\documentclass[twopage,12pt]{article}
\usepackage{amsmath,amssymb}
\allowdisplaybreaks[1]
\def\a{\alpha'}

\begin{document}

\begin{center}

{\Huge Higher-order string effective actions and off-shell $d=4$ supergravity}

\vskip 10mm

Filipe Moura

\vskip 4mm

Security and Quantum Information Group - Instituto de
Telecomunica\c c\~oes\\ Instituto Superior T\'ecnico, Departamento
de Matem\'atica \\Av. Rovisco Pais, 1049-001 Lisboa, Portugal

\vskip 2mm
and
\vskip 2mm

Centro de Matem\'atica da Universidade do Minho, Escola de Ci\^encias, \\Campus de Gualtar, 4710-057 Braga, Portugal.

\vskip 4mm

{\tt fmoura@math.ist.utl.pt}

\vskip 6mm

{\bf Abstract } \end{center}
\begin{quotation}\noindent

We start by a concise yet thorough revision of four-dimensional superspace
supergravity. We present curved superspace geometry, for arbitrary
${\cal N}$, including torsion, curvature and Bianchi identities.
We motivate the choice of torsion constraints.

We then consider the particular cases of ${\mathcal N}=1, 2.$ In
both cases we show how Poincar\'e supergravity can be obtained
from conformal supergravity. We see how to obtain the different
versions of the Poincar\'e off-shell theory, with distinct
compensating multiplets and sets of
auxiliary fields. For those versions of ${\mathcal N}=1, 2$
supergravities known as "old minimal", we present the solutions to
the Bianchi identities, their field content and we show how to
write superspace actions for these theories and their extensions
using chiral densities and chiral projectors.

As concrete applications, we study the supersymmetrization of the two possible
${\cal R}^4$ terms in $d=4$, which are both required as string corrections to
supergravity.

We conclude by discussing possible applications of these results to open problems on black holes in string theory.
\end{quotation}

\vfill
\eject
\newpage
\tableofcontents
%%%%%%%%%%%%%%%%%%%%%%%%%%%%%%%%%%%%%%%%%%%%%%%%%%%%%%%%%%%%%%%%%%%%%%
%%%%%%%%%%%%%%%%%%%%%%%%%%%%%%%%%%%%%%%%%%%%%%%%%%%%%%%%%%%%%%%%%%%%%%
\section{Introduction and plan}
Remarkable results have been achieved recently on black hole physics
in string theory, among which the microscopic interpretation of the
entropy and the attractor mechanism. Supersymmetry has played a crucial
role in these results.

Black holes can appear already at the supergravity level, when string
theories are compactified and (nonperturbative) \emph{p}-branes are
wrapped around nontrivial cycles of the compactification manifold. But
black holes can also be formed from elementary perturbative string
excitations; however, in this case the area of their horizons vanishes
at the supergravity level (these are called small black holes). In
order to prevent a naked singularity and get a finite horizon area,
one needs to consider the effect of higher-order string corrections
to supergravity. These terms appear in string theory effective actions
as $\a$ corrections, both at string tree level and higher string loops.
They also affect classical black holes, since they introduce corrections
to the supergravity equations of motion.

These are some reasons that motivate us to study higher-derivative
corrections to supergravity theories and their supersymmetrization. This is
what we do in the following, concentrating on theories in $d=4.$

We begin by reviewing four-dimensional superspace
supergravity. We present curved superspace geometry, for arbitrary
${\cal N}$, including torsion, curvature and Bianchi identities.
We motivate the choice of torsion constraints.

Next we move to the particular cases of ${\mathcal N}=1, 2.$ In
both cases we show how Poincar\'e supergravity can be obtained
from conformal supergravity by introducing a nonconformal
constraint. We see how different choices of this nonconformal
constraint lead to different versions of the Poincar\'e off-shell
theory, with distinct compensating multiplets and sets of
auxiliary fields. For those versions of ${\mathcal N}=1, 2$
supergravities known as "old minimal", we present the solutions to
the Bianchi identities, their field content and we show how to
write superspace actions for these theories and their extensions
using chiral densities and chiral projectors.

We then apply this formalism to the supersymmetrization of
higher-derivative terms in ${\mathcal N}=1, 2$
supergravities. As a concrete application, we study the supersymmetrization of
${\cal R}^4$ terms, which are required as string corrections to
those theories. We write down the ${\cal R}^4$ terms which appear in
the $\a^3$ type II and heterotic superstring effective actions. In
$d=4$ there are two of these terms. One of them is the square of
the Bel-Robinson tensor. We work out its ${\cal N}=1, 2$
supersymmetrizations, and we verify for both cases, with this
term, that some auxiliary fields can be eliminated and some cannot.
We identify these auxiliary fields and we interpret these results, which
should be generalized to other supersymmetric higher-derivative terms, in terms of the breaking of conformal
supergravity we discussed before.

The other ${\cal R}^4$ term cannot be directly supersymmetrized,
as in ${\mathcal N}=1$ it violates chirality. We show how to
circumvent this problem in ${\mathcal N}=1$ and we argue that it
should not be possible in ${\mathcal N}=8.$

We conclude by discussing possible applications of these results to open problems on black holes in string theory.

%%%%%%%%%%%%%%%%%%%%%%%%%%%%%%%%%%%%%%%%%%%%%%%%%%%%%%%%%%%%%%%%%%%%%%
%%%%%%%%%%%%%%%%%%%%%%%%%%%%%%%%%%%%%%%%%%%%%%%%%%%%%%%%%%%%%%%%%%%%%%

\section{Superspace geometry}
\label{sg}

%%%%%%%%%%%%%%%%%%%%%%%%%%%%%%%%%%%%%%%%%%%%%%%%%%%%%%%%%%%%%%%%%%%%%%
\subsection{Vielbein, connection, torsion and curvature}
Curved superspace is a manifold parameterized by the usual commuting
$x-$space coordinates $x^\mu$, plus a set of anticommuting spinorial
coordinates, their number depending on the number of space-time
dimensions in question and the number of supersymmetries ${\cal N}$.
In four dimensions, we have
\begin{equation}
z^\Pi =\left( x^\mu ,\theta^a_A, \theta^a_{\dot A} \right)
\end{equation}
with $\mu=0, \cdots, 3$, $A, \dot A =1, 2$, $a=1, \cdots, {\cal
N}$.

Symmetries that are manifest in curved superspace are general
supercoordinate transformations, with para\-meters $\xi^\Lambda$,
and tangent space (structure group) transformations, with parameters
$\Lambda^{MN}$. Curved superspace coordinates transform under
general reparameterizations as
\begin{equation}
z^\Pi \rightarrow z^{\prime \Pi }=z^\Pi +\xi^\Pi
\end{equation}
with $\xi^\Pi =\left( \xi^\mu ,\xi^a_A ,\xi^a_{\dot A}\right)$
defined as arbitrary functions of $z^\Pi$. $\xi^\mu$ corresponds to
the usual $x-$space diffeomorphisms (Einstein transformations);
$\xi^a_A ,\xi^a_{\dot A}$ are their supersymmetric extension: the
local supersymmetry transformations.

The main geometric objects of curved superspace are the
supervielbein $E_\Pi^M$ and the superconnection $\Omega_{\Lambda
N}^{\ \ \ \ P}$. These objects transform under general
supercoordinate transformations as
\begin{eqnarray}
\delta E_\Pi^{\ \ N}=\xi^\Lambda \partial _\Lambda E_\Pi^{\ N}+
\left(\partial_\Pi \xi^\Lambda \right) E_\Lambda^{\ N}, \\ \delta
\Omega _{\Lambda M}^{\ \ \ \ N}=\xi^\Pi \partial_\Pi
\Omega_{\Lambda M}^{\ \ \ N}+\left(\partial_\Lambda \xi^\Pi
\right) \Omega_{\Pi M}^{\ \ \ N}.
\end{eqnarray}

The supervielbein relates the curved indices to the tangent space
group ones, which we take to be $\mbox{SO}(1,3) \times {\mbox
U}({\cal N})$, with parameters $\Lambda_{MN}=\left( \Lambda_{mn},
\Lambda_{Bb Aa}, \Lambda_{\dot Bb \dot Aa}\right).$ These parameters
can still be decomposed in Lorentz and U($\cal N$) parts as
\begin{equation}
\Lambda_{Bb Aa}= \epsilon_{ba} \Lambda_{B A} + \epsilon_{BA}
\widetilde{\Lambda}_{ba}, \, \Lambda_{\dot Bb \dot Aa} =
\epsilon_{ba} \Lambda_{\dot B \dot A}+ \epsilon_{\dot B \dot A}
\widetilde{\Lambda}_{ba},
\end{equation}
satisfying
$$\Lambda_{B A}=\Lambda_{AB},\, \Lambda_{\dot B \dot A} =
\Lambda_{\dot A \dot B}, \, \Lambda_{A \dot A B \dot B}= 2
\varepsilon_{\dot A \dot B} \Lambda_{AB} +2 \varepsilon_{AB}
\Lambda_{\dot A \dot B} = - \Lambda_{B \dot B A \dot A}.$$ The
U($\cal N$) parameters can still be decomposed into SU($\cal N$) and
U(1) parts:
\begin{equation}
\widetilde{\Lambda}_{ba} = \Lambda_{ba}-\frac{1}{2} \epsilon_{ba}
\Lambda, \, \Lambda_a^{\ a}=0.
\end{equation}
About our choice of structure group, two remarks must be made.
Although the superconformal algebra is SU$\left(\left. 2,2\right|
\cal N \right)$, the superspace we have introduced is perfectly
adequate for the description of conformal supergravity. This is
because from the additional parameters of SU$\left(\left.
2,2\right| \cal N \right)$, special conformal boosts get absorbed
into general coordinate transformations, while Weyl (dilatations)
and special supersymmetry transformations will appear as extra
symmetries.

In principle we could have chosen some other structure group: if we
wanted a superspace formulation that mimicked the $x$-space
formulation of general relativity, the natural choice of structure
group would rather contain the orthosymplectic group OSp($\left.
1,3\right|4$) instead of the Lorentz group, but this would lead to
problems. Indeed, any superspace formulation of supergravity
requires the introduction of too many fields, through the
supervielbeins and the superconnections. The gauge invariances of
the theory allow one to eliminate some of the degrees of freedom,
but that is still not enough. In order to have a plausible theory,
in any superspace formulation one needs to put constraints on
covariant objects, so that the excess of fields (some of them of
spin exceeding two) can be eliminated. It can be shown (for
instance, in \cite{Van Nieuwenhuizen:1981ae}) that with such a
choice of tangent group one would not be able to put an adequate set
of constraints that could remove all the unwanted fields. The
largest group that allows that set of constraints is precisely the
one we took.

The supervielbein and superconnection transform under the
structure group as
\begin{eqnarray}
\delta E_\Pi^{\ \ N}&=&-E_\Pi^{\ \ M} \Lambda_M^{\ \ N}, \\ \delta
\Omega_{\Lambda M}^{\ \ \ \ N}&=&-\partial_\Lambda \Lambda_M^{\ \
N}+\Omega_{\Lambda M}^{\ \ \ S}\Lambda_S^{\ \ N}+\Omega_{\Lambda
R}^{\ \ \ N} \Lambda_M^{\ \ R}\left( -\right)^{\left( M+R\right)
\left( N+R\right)}. \label{deltastructure}
\end{eqnarray}

The superconnection is a structure algebra-valued (i.e. in the Lie
algebra of the structure group) object, which can of course also
be decomposed in its Lorentz and U($\cal N$) parts. Specifically,
the Lorentz part $\Omega_{\Lambda M}^{Lor \ N}$ is written as
\begin{equation}
\Omega_{\Lambda M}^{Lor \ N}=\left(\begin{array}{ccc}
\Omega_{\Lambda m}^{\ \ \ n} & 0 & 0 \\ 0 &
-\frac{1}{4}\Omega_\Lambda^{\ mn} \left( \sigma_{mn}\right)_B^{\
A} & 0 \\ 0 & 0 & \frac{1}{4} \Omega_\Lambda^{\
mn}\left(\sigma_{mn}\right)_{\dot B}^{\ \dot A} \end{array}
\right). \label{omegalor}
\end{equation}

Having the superconnection, we define a supercovariant derivative:
\begin{equation}
D_\Lambda =\partial_\Lambda +\frac{1}{2}\Omega_\Lambda^{\ \ MN}
J_{MN},\, \nabla_M=E_M^{\ \ \Lambda} D_\Lambda. \label{dcov}
\end{equation}
$J_{MN}$ are the generators of the structure group
($\left(\sigma_{mn}\right)_B^{\ A}, \left(\sigma_{mn}\right)_{\dot
B}^{\ \dot A}$ in the spinorial representation of the Lorentz
group). We define the (super)torsions $T_{MN}^{\ \ \ P}$ and
(super)curvatures $R_{MN}^{\ \ \ \ PQ}$ as
\begin{eqnarray}
T_{MN}^{\ \ \ \ R}&=&E_M^{\ \ \Lambda } \left(\partial_\Lambda
E_N^{\ \ \Pi} \right) E_\Pi^{\ R}+\Omega_{MN}^{\ \ \ \ R}- \left(
-\right)^{MN}\left(M \leftrightarrow N\right) \nonumber \\ &=&
E_M^{\ \ \Lambda} \left(D_\Lambda E_N^{\ \ \Pi} \right) E_\Pi^{\ R}
-
\left( -\right)^{MN}\left(M \leftrightarrow N\right), \label{tor} \\
R_{MN}^{\ \ \ \ RS}&=&E_M^{\ \ \Lambda }E_N^{\ \ \Pi }\left\{
\partial _\Lambda \Omega _\Pi^{\ RS}+\Omega _\Lambda^{\
RK}\Omega_{\Pi K}^{\ \ \ S}-\left( -\right)^{\Lambda \Pi }\left(
\Lambda \leftrightarrow \Pi \right) \right\} \nonumber
\\ &=&E_M^{\ \ \Lambda }E_N^{\ \ \Pi }\left\{ D_\Lambda
\Omega_\Pi^{\ RS}-\left( -\right)^{\Lambda \Pi } \left( \Lambda
\leftrightarrow \Pi \right) \right\}. \label{cur}
\end{eqnarray}
The curvatures are structure algebra-valued and, therefore, can
also be decomposed in their Lorentz and U($\cal N$) parts. Because
of (\ref{omegalor}), we have
\begin{eqnarray}
R_{MN C \dot C D \dot D} &=&2 \epsilon_{\dot C \dot D} R_{MNCD} +2
\epsilon_{CD} R_{MN \dot C \dot D}, \nonumber \\ R_{MNmn}&=&
-\frac{1}{2} \sigma_{mn}^{CD} R_{MNCD} -\frac{1}{2}
\sigma_{mn}^{\dot C \dot D} R_{MN \dot C \dot D}.
\end{eqnarray}

From the definitions (\ref{dcov}), (\ref{tor}) and (\ref{cur}) we
have, for the supercommutator of covariant derivatives,
\begin{equation}
\left[ \nabla_M,\nabla_N \right\} =T_{MN}^{\ \ \ \ R}\nabla_R
+\frac{1}{2}R_{MN}^{\ \ \ \ RS} J_{RS}. \label{supercom}
\end{equation}

Torsions and curvatures satisfy Bianchi identities. One of the
most important consequences of these identities is the fact that
the curvatures can be expressed completely in terms of the
torsions. This statement, known as Dragon's theorem
\cite{dragon79}, is also a consequence of the curvatures being
Lie-algebra valued. This fact has no place in general relativity,
where curvatures and torsions are independent, and one can
constrain the torsion to vanish leaving a nonvanishing curvature.
In superspace, the torsion is the main object determining the
geometry. The curvature Bianchi identity is therefore redundant;
all the information contained in it is also contained in the
torsion Bianchi identity, which is written as
\begin{eqnarray}
&-& \left( - \right)^{\left( M+N\right) R}\nabla _R T_{MN}^{\ \ \
\ F} +\left( -\right)^{\left( N+R\right) M}T_{NR}^{\ \ \ \
S}T_{SM}^{\ \ \ \ F} +\left( -\right)^{\left(N+R\right) M}
R_{NRM}^{\ \ \ \ \ \ F} \nonumber \\ &+& \left( -
\right)^{MN}\nabla _N T_{MR}^{\ \ \ \ F}-\left(- \right)^{NR}
T_{MR}^{\ \ \ \ S}T_{SN}^{\ \ \ \ F}- \left(
-\right)^{NR}R_{MRN}^{\ \ \ \ \ \ F}\ \nonumber \\ &-& \nabla_M
T_{NR}^{\ \ \ \ F} +T_{MN}^{\ \ \ \ S}T_{SR}^{\ \ \ \
F}+R_{MNR}^{\ \ \ \ \ \ F} =0.
\end{eqnarray}

%%%%%%%%%%%%%%%%%%%%%%%%%%%%%%%%%%%%%%%%%%%%%%%%%%%%%%%%%%%%%%%%%%%%%%
\subsection{Variational equations}
\label{ve} \indent

Arbitrary variations of supervielbein and superconnection are
given by \cite{wz781}
\begin{equation}
H_M^{\ \ N}=E_M^{\ \ \Lambda } \delta E_\Lambda^{\ \ N}, \,
\Phi_{MN}^{\ \ \ \ P} = E_M^{\ \ \Lambda } \delta \Omega_{\Lambda
N}^{\ \ \ \ P}. \label{phimn}
\end{equation}
From (\ref{tor}) and (\ref{phimn}), we derive the arbitrary
variation of the torsion:
\begin{eqnarray}
\delta T_{MN}^{\ \ \ \ R} &=& -H_M^{\ \ S} T_{SN}^{\ \ \ R}
+\left(-\right)^{MN} H_N^{\ \ S} T_{SM}^{\ \ \ R} +T_{MN}^{\ \ \ \
S} H_S^{\ \ R} \nonumber \\ &-&\nabla_M H_N^{\ \ R}
+\left(-\right)^{MN} \nabla_N H_M^{\ \ R} +\Phi_{MN}^{\ \ \ \
R}-\left(-\right)^{MN} \Phi_{NM}^{\ \ \ \ R}. \label{deltatmnr}
\end{eqnarray}

By matching (\ref{phimn}) to the variations under general
coordinate and structure group transformations, one can solve for
$H_M^{\ \ N}$ and $\Phi_{MN}^{\ \ \ \ P}$ in terms of the
transformation parameters, torsions and curvatures as
\begin{equation}
H_M^{\ N}= \xi^P T_{PM}^{\ \ \ N} + \nabla_M \xi^N +\Lambda_M^{\
N}, \, \Phi_{MN}^{\ \ \ \ P} =\xi^Q R_{QMN}^{\ \ \ \ \ \ P} -
\nabla_M \Lambda_N^{\ P}.
\end{equation}
Until a gauge for the general coordinate and structure group
transformations has not been fixed, any solution for $H_M^{\ \ N}$
and $\Phi_{MN}^{\ \ \ \ P}$ is valid up to the transformations
\begin{eqnarray}
\delta H_M^{\ \ \ N} &=& \nabla_M {\tilde \xi}^N- \xi^P T_{PM}^{\
\ \ \ N}, \,\, \delta \Phi_{MN}^{\ \ \ \ P} = {\tilde \xi}^Q
R_{QMN}^{\ \ \ \ \ \ P}, \label{gauge1} \\ \delta H_M^{\ \ N} &=&
{\widetilde \Lambda}_M^{\ \ N}, \,\, \delta \Phi_{MN}^{\ \ \ \ P}
= \nabla _M {\widetilde \Lambda}_N^{\ \ P}. \label{gauge2}
\end{eqnarray}
Even fixing those gauges does not fix all the degrees of freedom of
$H_M^{\ \ N}$ \cite{ht78,howe82, Muller:1988ux}. Namely,
$H=-\frac{1}{4} H_m^{\ m}$ remains an unconstrained superfield and
parametrizes the super-Weyl transformations, which include the
dilatations and the special supersymmetry transformations.

%%%%%%%%%%%%%%%%%%%%%%%%%%%%%%%%%%%%%%%%%%%%%%%%%%%%%%%%%%%%%%%%%%%%%%
\subsection{Choice of constraints}
\label{cc} \indent

As we previously mentioned, the superspace formulation of
supergravity requires the introduction of too many fields, some of
those having spins higher than 2. The only natural way to
eliminate the undesired fields and get only those belonging to an
irreducible representation of supersymmetry is to place
constraints in the theory. Since those constraints should be valid
in any frame of reference, they should be put only in covariant
objects; and since, as we saw, we can express the curvatures in
terms of the torsions, we choose to put the constraints in the
torsions. Therefore, using the gauge freedom from
(\ref{deltatmnr}), we analyze, from lower to upper dimensions,
which torsions we can constrain.

At dimension 0, we have the torsion parts $T_{A B}^{a b m}$,
$T_{A\dot B}^{a b m}$ and their complex conjugates. Considering
the flat superspace limit for $T_{A\dot B}^{a b m}$, we write
\begin{equation}
T_{A\dot B}^{a b m}=-2i\varepsilon^{a b} \sigma_{A \dot B}^m +
\tilde{T}_{A\dot B}^{a b m}.
\end{equation}
From (\ref{deltatmnr}), one finds \cite{muller862} that the only
parts of the torsion which cannot be absorbed by $H_n^{\ m}$,
$H_{a B}^{A b}$, $H_{a \dot B}^{A b}$ and their complex conjugates
are
\begin{equation}
\tilde{T}_{A\dot B C \dot C}^{a b} = \tilde{T}_{\underline{A}
\underline{\dot B} \underline{C} \underline{\dot C}}^{a b}, \,\,
T_{A B C \dot C}^{ab} = T_{\underline{A} \underline{B}
\underline{C} \dot C}^{\underline{a}\underline{b}},
\end{equation}
$\tilde{T}_{\underline{A} \underline{\dot B} \underline{C}
\underline{\dot C}}^{a b}$ being traceless in $a,b$. Since these
fields have spin greater than two and therefore it would be
impossible to describe any dynamics in their presence, we set them
to zero:
\begin{equation}
\tilde{T}_{\underline{A} \underline{\dot B} \underline{C}
\underline{\dot C}}^{a b} = 0, \,\, T_{\underline{A} \underline{B}
\underline{C} \dot C}^{\underline{a} \underline{b}} = 0.
\end{equation}
One must emphasize that these are the \emph{only} constraints
which have to be postulated (i.e. no other choice could be made to
these specific parts of the torsion). All the other constraints
are \emph{conventional}, which means they must exist, but other
choices could have been made. Conventional constraints correspond
to redefinitions of the supervielbein and superconnection.

We are then left with
\begin{equation}
T_{A\dot B}^{a b m}=-2i\varepsilon^{a b} \sigma_{A \dot B}^m, \,\,
T_{A B}^{a b m} =0.
\end{equation}

As we will see, in ${\cal N}=1,2$ theories the constraint $T_{A
B}^{a b m}=0$ has a geometrical meaning, and will be called
"representation preserving". The constraint in $T_{A\dot B}^{a b
m}$ is just conventional.

At dimension $\frac{1}{2}$, it can be shown that, by adequate
choices of the suitable parts of $H_N^{\ M}$ and $\Phi_{MN}^{\ \ \
\ P}$ \cite{muller862}, we may set
\begin{equation}
T_{A a \dot B b \dot C c} = 0, \,\, T_{A a B b C c} = 0, \,\,
T_{A}^{a m n} = 0.
\end{equation}
At dimension 1, an appropriate redefinition of the Lorentz
connection through an adequate choice of $\Phi_{mn}^{\ \ \ p}$
gives the usual constraint in Riemannian geometry
\begin{equation}
T_{m n}^{\ \ \ p}=0.
\end{equation}
Also, an adequate choice of $\Phi_{ma}^{\ \ \ b}$ allow us to
constrain $R_{C cab}^{c \dot C}$, and to have
\begin{equation}
T_{C \dot C B}^{\ \ \ \ bCa}=\beta T_{C \dot C \ \ B}^{\ \ \ Cba}.
\end{equation}
This constraint establishes an identity between two a priori
different superfields. The numerical parameter $\beta$ depends on
the choice that was made for $R_{C cab}^{c \dot C}$, but it will
have no impact on the theory, since this is a conventional
constraint.

The Bianchi identities are valid, no matter which constraints we
have. But once some of the torsions are constrained, the Bianchi
identities become equations for the unconstrained torsions and
curvatures. These equations are not independent, and need to be
solved systematically. This has been achieved, in conformal
supergravity, for arbitrary ${\cal N}$ \cite{howe82}. One can
conclude that off-shell conformal supergravity exists and is
consistent for ${\cal N} \leq 4$. For ${\cal N} \geq 6$, an
off-shell theory is not consistent \cite{howe82, Muller:1988ux}.
That does not rule out on-shell theories, but those have not been
found. For ${\cal N}=5$ nothing has been concluded. Thus for ${\cal
N} >4$ the situation is rather unclear. We will only review the
${\cal N}=1,2$ cases, because those are the ones we will need. For a
more complete discussion the reader is referred to
\cite{Muller:1988ux}.

In ${\cal N}=1, 2$ one can put chirality constraints in
superfields. An antichiral superfield $\Phi_{\cdots}$ satisfies
\begin{equation}
\nabla_A^a \Phi_{\cdots} = 0 \label{ac}
\end{equation}
(the hermitean conjugated equation defines a chiral superfield).
This constraint on the superfield must be compatible with the
solution to the Bianchi identities; an integrability condition
must be verified (that is why general chiral superfields only
exist for ${\cal N}=1, 2$, as we will see; for other values of
${\cal N}$, a chirality condition may result only from the
solution to the Bianchi identities, in the superfields introduced
in this process).

${\cal N}=1,2$ Poincar\'e supergravities can be obtained from the
corresponding conformal theories by consistent couplings to
compensating multiplets that break superconformal invariance and
local U($\cal N$). There are different possible choices of
compensating multiplets, leading to different formulations of the
Poincar\'e theory. What is special about these theories is the
existence of a completely off-shell formalism. This means that,
for each of these theories, a complete set of auxiliary fields is
known (actually, there exist three known choices for each theory).
In superspace this means that, after imposing constraints on the
torsions, we can completely solve the Bianchi identities without
using the field equations \cite{howe82,gwz79}, and there is a
perfect identification between the superspace and $x$-space
descriptions. We will review how is this achieved for the "old
minimal" ${\cal N}=1,2$ cases.

%%%%%%%%%%%%%%%%%%%%%%%%%%%%%%%%%%%%%%%%%%%%%%%%%%%%%%%%%%%%%%%%%%%%%%
%%%%%%%%%%%%%%%%%%%%%%%%%%%%%%%%%%%%%%%%%%%%%%%%%%%%%%%%%%%%%%%%%%%%%%
\section{${\cal N}=1$ supergravity in superspace}

\subsection{${\cal N}=1$ superspace geometry and constraints}
\label{n=1}

${\cal N}=1$ superspace geometry is a simpler particular case of
the general ${\cal N}$ case we saw in the previous section.
Namely, the internal group indices $a, b, \cdots$ do not exist.
The structure group is at most $\mbox{SO}(1,3) \times {\mbox
U}(1)$ (in the Poincar\'e theory we will consider, it is actually
just the Lorentz group). To write any U(${\cal N}$) valued formula
in the ${\cal N}=1$ case, one simply has to decompose that formula
under U(${\cal N}$) and take simply the group singlets.

Specific to ${\cal N}=1, 2$ are the representation-preserving
constraints, required by the above mentioned integrability
condition for the existence of antichiral superfields, defined by
(\ref{ac}). For ${\cal N}=1$, these constraints are the following:
\begin{equation}
T_{AB}^{\ \ \ \dot C} =0, \, T_{AB}^{\ \ \ m}=0.
\end{equation}

Conventional constraints allow us to express the superconnection in
terms of the supervielbein. Namely, the constraint $T_{m n p}=0$
allow us to solve for the bosonic connection $\Omega_{mn}^{\ \ \
p}$, exactly as in general relativity. Constraints $T_{AB}^{\ \ \ C}
=0$ allow us to solve for $\Omega_{AB}^{\ \ \ C}$, and $T_{A
\underline{\dot B}}^{\ \ \ \underline{\dot C}} =0$, for
$\Omega_{A\dot B}^{\ \ \ \dot C}$. But in ${\cal N}=1$ supergravity
one can even go further, and solve for the supervielbein parts with
bosonic tangent indices $E_n^{\ \ \Pi}$ in terms of the other parts
of the supervielbein. The conventional constraints that allow for
that are $T_{A\dot B}^{\ \ \ m}=-2i\sigma_{A \dot B}^m$, $T_{A\dot
B}^{\ \ \ \dot C} -\frac{1}{4} T_A^{\ mn}
\left(\sigma_{mn}\right)_{\dot B}^{\ \dot C}=0$.

In section \ref{cc}, we required a stronger constraint, which in
${\cal N}=1$ language is written as $T_A^{\ mn}=0$. We can still
require that as a conventional constraint, if we take for structure
group $\mbox{SO}(1,3) \times {\mbox U}(1)$. In the formulations in
which U(1) is not gauged, only the constraints above are taken for
the conformal theory, but an extra constraint will be necessary in
order to obtain the Poincar\'e theory. We will analyze the possible
cases next.
%%%%%%%%%%%%%%%%%%%%%%%%%%%%%%%%%%%%%%%%%%%%%%%%%%%%%%%%%%%%%%%%%%%%%%
%%%%%%%%%%%%%%%%%%%%%%%%%%%%%%%%%%%%%%%%%%%%%%%%%%%%%%%%%%%%%%%%%%%%%%
\subsection{From conformal to Poincar\'e supergravity}

To obtain ${\cal N}=1$ Poincar\'e supergravity from conformal
supergravity, we must adopt contraints which do not preserve the
superconformal invariance. However, we must not break all
superconformal invariance, since that would be equivalent to fixing
all the superconformal gauges, and we would be left only with the
fields which are inert under superconformal gauge choices, i.e. the
fields of the Weyl multiplet $e_\mu^m$, $\psi_\mu^A$ and $A_\mu$. As
we will see, this will be the case either with gauged or with
ungauged U(1).

%%%%%%%%%%%%%%%%%%%%%%%%%%%%%%%%%%%%%%%%%%%%%%%%%%%%%%%%%%%%%%%%%%%%%%
\subsubsection{Ungauged U(1)}

To determine the nonconformal constraints, we must first determine
the transformation properties of the supervielbeins and
superconnections.

In Lorentz superspace, the super Weyl parameter $L$ is complex. We
define
\begin{equation}
E_A^{\prime \ \Pi}=e^L E_A^{\ \ \Pi}, \, \, E_{\dot A}^{\prime \
\Pi}=e^{\overline{L}} E_{\dot A}^{\ \ \Pi}. \label{rescvil}
\end{equation}

Since, with our choice of constraints, supervielbeins and
superconnections can all be expressed in terms of the spinor
vielbeins, we only need these transformation properties;
conventional constraints are valid for any set of vielbeins and,
therefore, they are automatically satisfied when one replaces
$E_A^{\ \ \Pi}$, $E_{\dot A}^{\ \ \Pi}$ by their rescaled values.
Then it can be proven \cite{bk} that the representation preserving
constraints are invariant under (\ref{rescvil}). If these
constraints were not invariant under Weyl transformations, then
chiral multiplets could not exist in the background of conformal
supergravity.

A complex scalar superfield can be decomposed in local superspace
into chiral and linear parts. After breaking part of the super-Weyl
group, the parameters $L, \overline{L}$ will be restricted such that
a linear combination of them will be either chiral or linear. In the
first case, one needs a dimension $\frac{1}{2}$ constraint; in ths
second, one of dimension 1. The only left unconstrained objects of
dimension $\frac{1}{2}$ and 1 are, respectively, the torsion
component $T_{Am}^{\ \ \ m}$ and the superfield $R=R_{AB}^{\ \ \
AB}$. These superfields transform under the super-Weyl group as
$\left(\nabla^2 = \nabla^A \nabla_A\right)$ \cite{bk, Gates:1983nr}
\begin{equation}
T_{Am}^{\prime \ \ m}= e^L \left(T_{Am}^{\ \ \ m} +2 \nabla_A
\left(2L +\overline{L} \right) \right), \, R^\prime=3 \left(\nabla^2
+\frac{1}{3} R \right) e^{2L}.
\end{equation}
We can break the super Weyl invariance by imposing as a constraint
\begin{equation}
T_{Am}^{\ \ \ m}=0.
\end{equation}
For that to be consistent, we must impose that $2L+\overline{L}$ is
antichiral:
\begin{equation}
\nabla_A \left(2L+\overline{L} \right)=0. \label{ht}
\end{equation}
What is left from the super-Weyl group is the so-called Howe-Tucker
group \cite{ht78}: the supervielbeins transforming as in
(\ref{rescvil}), with $L,\overline{L}$ satisfying (\ref{ht}).

This constraint leads to the "old minimal" formulation of ${\cal
N}=1$ Poincar\'e supergravity \cite{sw78,fvn78}. To the Weyl
multiplet of conformal supergravity we are adding a compensating
chiral multiplet with 8+8 components.

Another possibility to break the super Weyl invariance is to set
the constraint $R=0$; the remaining super Weyl invariance contains
a parameter $L$ that now is an antilinear superfield: $\nabla^2
L=0.$ This constraint leads to the nonminimal formulation of
${\cal N}=1$ Poincar\'e supergravity \cite{s79}. To the Weyl
multiplet of conformal supergravity we are adding a compensating
linear multiplet having 12+12 components. This way, we have
fermionic auxiliary fields.

Both constraints can be generalized. On dimensional grounds, the
most general nonconformal constraint one may take is given by
\cite{bk,Gates:1983nr}
\begin{equation}
C=-\frac{1}{3} R +\frac{n+1}{3n+1} \nabla^A T_{Am}^{\ \ \ m}
-\left( \frac{n+1}{3n+1}\right)^2 T^{Am}_{\ \ \ m} T_{An}^{\ \ \
n}=0. \label{defn}
\end{equation}
$n$ is a numerical parameter. This constraint transforms, for small
$L$, as
\begin{equation}
\delta C = 2LC -2\left(\nabla^2-2 \frac{n+1}{3n+1} T^{Am}_{\ \ \
m} \nabla_A \right) \left(L- \frac{n+1}{3n+1}
\left(2L+\overline{L}\right)\right).
\end{equation}
For a generic choice of $n$, the constraint $R=0$ is necessary and
we have a nonminimal formulation. Taking $n=-\frac{1}{3}$
corresponds to the "old minimal" formulation we saw.

Another interesting case occurs by taking $n=0$: only
$L+\overline{L}$ appears in $\delta C$, such that the (axial) U(1)
local gauge invariance, which we did not include in the structure
group, is actually conserved, with parameter $L-\overline{L}$. This
corresponds to the "new minimal" (also known as "axial") formulation
of ${\cal N}=1$ Poincar\'e supergravity \cite{sw81}, in which one
introduces a compensating tensor multiplet having 8+8 components.

Whichever constraint we choose, the irreducible parameter invariance
of the resulting geometry corresponds to the compensating multiplet.
This invariance allows for redefinition of torsions and, after
gauge-fixing, for the fields of the compensating multiplet to appear
in the final theory, with the original symmetry completely broken.
These are very generical features, which we will also meet in the
formulation of the ${\cal N}=2$ theory.

%%%%%%%%%%%%%%%%%%%%%%%%%%%%%%%%%%%%%%%%%%%%%%%%%%%%%%%%%%%%%%%%%%%%%%
\subsubsection{Gauged U(1)}

Let's now start from a $\mbox{SO}(1,3) \times {\mbox U}(1)$
superspace. From (\ref{deltastructure}), the fermionic part of the
U(1) connection transforms under U(1) as ($A$ is a "flat" index):
\begin{equation}
\delta \Omega_A=-\nabla_A \Lambda - \Lambda \Omega_A
\end{equation}
while, from (\ref{phimn}), under a general transformation we have
\begin{equation}
\delta \Omega_A=\Phi_A - H_A^{\ M} \Omega_M.
\end{equation}
In U(1) superspace, after fixing the constraints it can be shown
\cite{howe82} that one has $H_{AB} = \frac{1}{2} \varepsilon_{AB}
H$, $\Phi_A= \frac{3}{2} \nabla_A H$, $H=-\frac{1}{4} H_m^{\ m}$
being an unconstrained superfield defined in section \ref{ve}
which parametrizes the super-Weyl transformations. Overall,
$\Omega_A$ transforms as
\begin{equation}
\delta \Omega_A=\nabla_A \left(\frac{3}{2} H -\Lambda \right) +
\left(\frac{1}{2} H -\Lambda \right) \Omega_A.
\end{equation}
From this transformation law, by setting the constraint
$\Omega_A=0,$ we see that we break the superconformal and local
U(1) symmetries and restrict the combination $\frac{3}{2} H
-\Lambda$ to a compensating chiral multiplet. This is the "old
minimal" formulation of ${\cal N}=1$ Poincar\'e supergravity
\cite{sw78,fvn78}. Other formulations have a treatment similar to
the ungauged U(1) case. From now on, by ${\cal N}=1$ Poincar\'e
supergravity we always mean the "old minimal" formulation with
$n=-1/3$.
%%%%%%%%%%%%%%%%%%%%%%%%%%%%%%%%%%%%%%%%%%%%%%%%%%%%%%%%%%%%%%%%%%%%%%
%%%%%%%%%%%%%%%%%%%%%%%%%%%%%%%%%%%%%%%%%%%%%%%%%%%%%%%%%%%%%%%%%%%%%%
\subsection{The chiral compensator and the chiral measure}

The superspace approach we have discussed has the inconvenience of
involving a large number of fields and a large symmetry group.
This way, one must put constraints and choose a particular gauge
to establish the compatibility to the $x$-space theory (see
section \ref{a1}). There is an approach which uses from the
beginning fewer fields and a smaller symmetry group (holomorphic
general coordinate transformations) \cite{Van
Nieuwenhuizen:1981ae, bk, Gates:1983nr, gs80, os78, lr79}. In this
approach we take two chiral superspaces with complex coordinates
$\left(y^\mu, \theta \right), \, \, \left(\overline{y^\mu},
\overline{\theta} \right),$ which are related by complex
conjugation. In four-component spinor notation, $\theta =
\frac{1}{2} \left(1+\gamma_5 \right) \Theta, \, \,
\overline{\theta} = \frac{1}{2} \left(1-\gamma_5 \right) \Theta.$
One also has
\begin{equation}
\frac{1}{2} \left(y^\mu + \overline{y^\mu} \right)= x^\mu, \,\,
y^\mu - \overline{y^\mu} = 2i H^\mu\left(x, \Theta \right).
\end{equation}
This way, the imaginary part of the coordinates $y^\mu,
\overline{y^\mu}$ is interpreted as an axial vector superfield,
while the real part is identified with real spacetime. One has
then in the combined $8+4$ dimensional space $\left(y^\mu,
\overline{y^\mu}, \theta, \overline{\theta} \right)$ a $4+4$
dimensional hypersurface defined by $y^\mu - \overline{y^\mu}= 2i
H^\mu\left(y^\mu + \overline{y^\mu}, \theta, \overline{\theta}
\right).$ When one shifts points by a coordinate transformation,
the hypersurface itself is deformed in such a way that the new
points lie on the new hypersurface. These hypersurfaces, each
characterized by the superfield $H^\mu\left(y^\mu +
\overline{y^\mu}, \theta, \overline{\theta} \right)$, represent
each a real superspace like the one we have been working with.

The holomorphic coordinate transformations form a supergroup. If
one puts no further restriction on their parameters, one is led to
conformal supergravity. However, Poincar\'e supergravity is
described by the very natural subgroup of unimodular holomorphic
transformations, which satisfy
\begin{equation}
\mbox{sdet} \frac{\partial \left(y^{\mu \prime}, \theta^\prime
\right)}{\partial\left(y^\mu, \theta \right)}=1. \label{uni}
\end{equation}
One can take Poincar\'e supergravity is a gauge theory with the
gravitational superfield $H^\mu\left(x, \theta \right)$ as a
dynamical object and the supergroup of holomorphic coordinate
transformations being the gauge group \cite{os78}. But one can
also remove the constraint (\ref{uni}) and handle arbitrary
holomorphic transformations at the cost of the appearance of a
compensating superfield. In the "old minimal" $n=-1/3$ theory,
this superfield, which we define as $\varphi\left(y^\mu, \theta
\right)$, is holomorphic and is called the chiral compensator. It
transforms as \cite{bk,Gates:1983nr}
\begin{equation}
\varphi\left(y^\mu, \theta \right) = \left[\mbox{sdet}
\frac{\partial \left( y^{\mu \prime}, \theta^\prime
\right)}{\partial\left(y^\mu, \theta \right)}
\right]^{\frac{1}{3}} \varphi\left(y^{\mu \prime}, \theta^\prime
\right).
\end{equation}
One can then find a coordinate system in which
$\varphi\left(y^\mu, \theta \right) = 1.$ Clearly, all the
holomorphic coordinate transformations preserving this gauge are
unimodular; this way, we recover the gauge group of Poincar\'e
supergravity. Poincar\'e supergravity is then a theory of two
dynamical objects \cite{gs80} - the gravitational superfield
$H^\mu\left(x^\mu, \theta, \overline{\theta} \right)$ and the
chiral compensator $\varphi\left(y^\mu, \theta \right)$ -,
transforming under the supergroup of holomorphic coordinate
transformations, and defined in real superspaces, given by the
hypersurfaces above.

The chiral compensator allows us to define an invariant chiral
measure in superspace. Since
\begin{equation}
d^4 y d^2 \theta = \mbox{sdet} \frac{\partial\left(y^\mu, \theta
\right)}{\partial \left(y^{\mu \prime}, \theta^\prime \right)} d^4
y^{\prime} d^2\theta^\prime,
\end{equation}
we have
\begin{equation}
\varphi^3\left(y^\mu, \theta \right) d^4 y d^2 \theta =
\varphi^{\prime 3} \left(y^{\mu \prime}, \theta^\prime \right) d^4
y^{\prime} d^2\theta^\prime.
\end{equation}
We define then the chiral density \cite{wz781, bk, Gates:1983nr,
lr79} as
\begin{equation}
\epsilon=\varphi^3. \label{phi3}
\end{equation}
From the transformation law of $\varphi$, one can see that
$\epsilon$ transforms under supercoordinate transformations with
parameters $\xi^\Lambda$ as
\begin{equation}
\delta \epsilon = - \partial_\Lambda \left(\epsilon \xi^\Lambda
(-)^\Lambda \right). \label{deltaepsilon}
\end{equation}

Instead of choosing the gauge $\varphi\left(y^\mu, \theta \right)
= 1$, it is more convenient to choose a Wess-Zumino gauge for
$H^\mu$, in which this superfield is expressed only in terms of
the physical and auxiliary fields from the supergravity multiplet.
After fixing the remaining gauge freedom, the same is valid for
$\epsilon$.

%%%%%%%%%%%%%%%%%%%%%%%%%%%%%%%%%%%%%%%%%%%%%%%%%%%%%%%%%%%%%%%%%%%%%%
%%%%%%%%%%%%%%%%%%%%%%%%%%%%%%%%%%%%%%%%%%%%%%%%%%%%%%%%%%%%%%%%%%%%%%
\subsection{Solution to the Bianchi identities in "old minimal" ${\cal N}=1$
Poincar\'e supergravity}

The full off-shell solution to the Bianchi identities, given the
representation-preserving and conventional constraints in section
\ref{n=1} and the nonconformal constraint $T_{Am}^{\ \ \ m}=0$, is
standard textbook material which we do not include here
\cite{gwz79,wessbagger}. The results, in our conventions, may be
seen in \cite{Moura:2001xx}. It can be shown that, as a result of
$T_{Am}^{\ \ \ m}=0$ and the conventional constraint $T_{A\dot B}^{\
\ \ \dot C} -\frac{1}{4} T_A^{\ mn} \left(\sigma_{mn}\right)_{\dot
B}^{\ \dot C}=0$, one actually has simply $T_{Am}^{\ \ \ m}=0$ and
actually recovers the conventional constraint from the approach with
gauged U(1).

The off-shell solutions are described in terms of the supergravity
superfields $R=R_{AB}^{\ \ \ AB}$, $G_m$, $W_{ABC}$, their complex
conjugates and their covariant derivatives. $R$ and $W_{\dot A\dot
B\dot C}$ are antichiral:
\begin{equation}
\nabla^A R=0, \,\, \nabla_A W_{\dot A\dot B\dot C}=0.
\end{equation}

In ${\cal N}=1$, chiral superfields may exist with any number of
undotted indices (but no dotted indices). Chiral projectors exist;
when acting with them on any superfield with only undotted indices,
a chiral superfield always results. For scalar superfields the
antichiral projector is given by $\left(\nabla^2 +\frac{1}{3} R
\right).$

The torsion constraints imply the following off-shell differential
relations (not field equations) between the ${\cal N}=1$
supergravity superfields:
\begin{eqnarray}
\nabla^A G_{A\dot B} &=& \frac{1}{24} \nabla_{\dot B} R,
\label{diffg} \\ \nabla^A W_{ABC} &=& i \left( \nabla_{B \dot A}
G_C^{\ \ \dot A} +\nabla_{C \dot A} G_B^{\ \ \dot A} \right),
\label{diffw}
\end{eqnarray}
which, together with the torsion conventional constraints, imply the
relation
\begin{equation}
\nabla^2 \overline{R} - \overline{\nabla}^2 R = 96 i \nabla^n G_n.
\label{wb203}
\end{equation}

%%%%%%%%%%%%%%%%%%%%%%%%%%%%%%%%%%%%%%%%%%%%%%%%%%%%%%%%%%%%%%%%%%%%%%
%%%%%%%%%%%%%%%%%%%%%%%%%%%%%%%%%%%%%%%%%%%%%%%%%%%%%%%%%%%%%%%%%%%%%%

\subsection{From superspace to $x$-space} \label{a1}

Another special feature of pure ${\cal N}=1$ four-dimensional
supergravity is that its action in superspace is known. It is
written as the integral, over the whole superspace, of the
superdeterminant of the supervielbein \cite{Van
Nieuwenhuizen:1981ae, wz781}:

\begin{equation}
{\cal L}_{SG} =\frac{1}{2 \kappa^2} \int E\,d^4\theta\, ,
E=\mbox{sdet}E_\Lambda^{\ \ M}. \label{pure}
\end{equation}
On dimensional grounds, this is the only possible action. The
$\frac{1}{2 \kappa^2}$ factor is necessary to reproduce the
$x$-space results; in principle, one could multiply this action by
any dimension zero unconstrained scalar, but that object does not
exist. In this action, and in actions written as $d^4\theta$
integrals, the indices of the $\theta$-variables are curved, i.e.
they vary under Einstein transformations.

In order to determine the component expansion of this action, the
best is certainly not to determine directly all the components of
$E$, but rather to determine the component expansion of the
supergravity superfields. For that, we use the method of gauge
completion \cite{wessbagger,wz782}. The basic idea behind it is to
relate in superspace some superfields and superparameters at
$\theta=0$ (which we symbolically denote with a vertical bar on
the right) with some $x$ space quantities, and then to require
compatibility between the $x$ space and superspace transformation
rules \cite{sw78,fvn78}.

We make the following identification for the supervielbeins at
$\theta=0$ $\left. E_\Pi^{\ N }\right|$:
\begin{equation}
\left. E_\Pi^{\ \ N}\right| =\left[ \begin{array}{ccc} e_\mu^{\ \ m}
&
\frac{1}{2}\psi_\mu^{\ \ A } & \frac{1}{2}\psi_\mu^{\ \ \dot A }\\
0 & \delta_B^{\ \ A} & 0 \\ 0 & 0 & \delta_{\dot B}^{\ \ \dot A}
\end{array} \right]. \label{vielbeinx}
\end{equation}

In the same way, we gauge the fermionic superconnection at order
$\theta=0$ to zero and we can set its bosonic part equal to the
usual spin connection:
\begin{equation}
\left. \Omega_{\mu m}^{\ \ \ n} \right| = \omega_{\mu m}^{\ \ \ n}
\left( e, \psi\right), \left. \Omega_{A m}^{\ \ \ n}\right|, \left.
\Omega_{\dot A m}^{\ \ \ n}\right| =0. \label{connectionx}
\end{equation}
The spin connection is given, in ${\cal N}=1$ supergravity, by
\begin{eqnarray}
\omega_{\mu m}^{\ \ \ n}\left( e, \psi\right)&=& \omega_{\mu m}^{\ \
\ n}(e) -\frac{i}{4} \kappa^2 \left(\psi_{\mu A} \sigma_m^{A \dot A}
\psi^n_{\dot A} -\psi_{\mu A} \sigma^{n A \dot A} \psi_{m \dot A}
+\psi_{m A} \sigma_\mu^{A \dot A} \psi^n_{\dot A} \right. \nonumber \\
&+& \left. \psi_{\mu \dot A} \sigma_m^{A \dot A} \psi^n_A -\psi_{\mu
\dot A} \sigma^{n A \dot A} \psi_{m A} +\psi_{m \dot A}
\sigma_\mu^{A \dot A} \psi^n_A \right).
\end{eqnarray}
$\omega_{\mu m}^{\ \ \ n}(e)$ is the connection from general
relativity. We also identify, at the same order $\theta=0$, the
superspace vector covariant derivative (with an Einstein indice)
with the curved space covariant derivative: $\left. D_\mu \right| =
{\cal D}_\mu.$ These gauge choices are all preserved by supergravity
transformations.

As a careful analysis using the solution to the Bianchi identities
and the off-shell relations among the supergravity superfields
$\overline{R}, G_n, W_{ABC}$ shows, the component field content of
these superfields is all known once we know $$\left. \overline{R}
\right|, \left. \nabla_A \overline{R}\right|, \left. \nabla^2
\overline{R}\right|, \left. G_{A \dot A}\right|, \left.
\nabla_{\underline{A}}G_{\underline{B} \dot A}\right|, \left.
\nabla_{\underline{\dot A}} \nabla_{\underline{A}} G_{\underline{B}
\underline{\dot B}}\right|, \left. W_{ABC} \right|, \left. \nabla
_{\underline{D}}W_{\underline{ABC}}\right|.$$

All the other components and higher derivatives of $\overline{R},
G_{A \dot A}, W_{ABC}$ can be written as functions of these
previous ones. In order to determine the "basic" components, first
we solve for superspace torsions and curvatures in terms of
supervielbeins and superconnections using (\ref{vielbeinx}) and
(\ref{connectionx}); then we identify them with the off-shell
solution to the Bianchi identities \cite{bk, wessbagger, wz782}
\footnote{ $\psi_{\mu \nu}^B= {\cal D}_\mu \psi_\nu^B- {\cal
D}_\nu \psi_\mu^B$ is the gravitino curl.}: $$\left.
\overline{R}\right| = 4 \left( M+iN \right), \, \left. G_{A\dot
A}\right| =\frac{1}{3} A_{A\dot A}, \, \left. W_{ABC} \right|
=-\frac{1}{4} \psi_{\underline{A} \ \ \underline{B} \dot C
\underline{C}}^{\ \ \dot C} -\frac{i}{4} A_{\underline{A}\ }^{\ \
\dot C} \psi_{\underline{B} \dot C \underline{C}}.$$

$\left. R \right|$ and $\left. G_m \right|$ are auxiliary fields.
$A_\mu$, a gauge field in conformal supergravity, is an auxiliary
field in Poincar\'e supergravity. The (anti)chirality condition on
$R, \overline{R}$ implies their $\theta=0$ components (resp. the
auxiliary fields $M-iN, M+iN$) lie in antichiral/chiral multiplets
(the compensating multiplets); (\ref{diffg}) shows the spin-1/2
parts of the gravitino lie on the same multiplets (because, as we
will see in the next section, $\nabla_A G_{B \dot B}$, at
$\theta=0$, is the gravitino curl) and, according to
(\ref{wb203}), so does $\partial^\mu A_\mu$.

$\left. \nabla_A \overline{R}\right|, \left.
\nabla_{\underline{A}}G_{\underline{B} \dot A}\right|$ also come
straightforwardly from comparison to the solution to the Bianchi
identities \cite{bk, Moura:2001xx}. Finding $\left. \nabla^2
\overline{R}\right|, \left. \nabla_{\underline{\dot A}}
\nabla_{\underline{A}} G_{\underline{B} \underline{\dot
B}}\right|, \left. \nabla
_{\underline{D}}W_{\underline{ABC}}\right|$ is a bit more
involved: one must identify the (super)curvature $R_{\mu \nu}^{\ \
\ mn}$ with the $x$-space curvature ${\cal R}_{\mu \nu}^{\ \ \
mn}$, multiply by the inverse supervielbeins $E_M^{\ \mu} E_N^{\
\nu}$, identify with the solution to the Bianchi identities for
$R_{MN}$ and extract the field contents by convenient index
manipulation. The field content of these components will include
the Riemann tensor in one of its irreducible components,
respectively the Ricci scalar, the Ricci tensor and the selfdual
Weyl tensor (${\cal W}_{ABCD}:=-\frac{1}{8}{\cal W}^+_{\mu \nu
\rho \sigma} \sigma^{\mu \nu}_{\underline{AB}} \sigma^{\rho
\sigma}_{\underline{CD}},$${\cal W}^{\mp}_{\mu \nu \rho \sigma}
:=\frac{1}{2} \left({\cal W}_{\mu \nu \rho \sigma} \pm \frac{i}{2}
\varepsilon_{\mu \nu}^{\ \ \ \lambda \tau} {\cal W}_{\lambda \tau
\rho \sigma} \right)$). The full results are derived in
\cite{Moura:2001xx}; at the linearized level,
\begin{eqnarray}
\left. \nabla^2 \overline{R} \right| = -8 {\cal R} &+& \ldots, \,
\left. \nabla_{\underline{A}} \nabla_{\underline{\dot A}}
G_{\underline{B} \underline{\dot B}}\right| = -\frac{1}{2}
\sigma^\mu_{\underline{A} \underline{\dot A}}
\sigma^\nu_{\underline{B} \underline{\dot B}} \left( {\cal R}_{\mu
\nu} -\frac{1}{4} g_{\mu \nu} {\cal R} \right) + \ldots, \nonumber
\\ \left. \nabla_{\underline{A}} W_{\underline{BCD}} \right| &=&
-\frac{1}{8} {\cal W}^+_{\mu \nu \rho \sigma} \sigma^{\mu
\nu}_{\underline{AB}} \sigma^{\rho \sigma}_{\underline{CD}} +
\ldots, \left. \nabla^2 W^2 \right| = -2{\cal W}_+^2 + \ldots,
\label{dw0}
\\ \left. \nabla_{\underline{\dot A}} W_{\underline{\dot B \dot C \dot D}} \right|
&=& -\frac{1}{8} {\cal W}^-_{\mu \nu \rho \sigma} \sigma^{\mu
\nu}_{\underline{\dot A \dot B}} \sigma^{\rho
\sigma}_{\underline{\dot C \dot D}} + \ldots, \left.
\overline{\nabla}^2 \overline{W}^2 \right| = -2{\cal W}_-^2 +
\ldots \label{dw0dot}
\end{eqnarray}

Knowing these components, we can compute, in $x$-space, any action
which involves the supergravity multiplet. In order to do that, we
need to know how to convert superspace actions to $x$-space
actions.

Consider the coupling of a real scalar superfield to supergravity
given by
\begin{eqnarray}
{\cal L}&=&\frac{1}{2 \kappa^2} \int E \Phi \label{actiong}
d^4\theta = \frac{3}{4 \kappa^2} \int \left[\frac{E}{\overline
R}\left( \overline{\nabla}^2 +\frac{1}{3} {\overline R} \right)
+\frac{E}{R} \left( \nabla^2 +\frac{1}{3} R \right) \right] \Phi
d^4\theta \nonumber \\ &=& \frac{3}{4 \kappa^2} \int \left(
-\frac{1}{4} \frac{\overline{D}^2 E} {\overline{R}} \right) \left[
\left(\overline{\nabla}^2 +\frac{1}{3} {\overline R} \right) \Phi
\right] d^2\theta + \mathrm{h.c.}. \label{dnabla}
\end{eqnarray}
$D_A=\left(E^{-1}\right)_A^{\ M} \nabla_M$ is the superspace
covariant derivative with an Einstein index. In the previous
equation, the operator $d^2\overline{ \theta}=-\frac{1}{4}
\overline{D}^2$ should apply to all the integrand, and not only to
$E$. But, knowing that we can choose the gauge (\ref{vielbeinx}),
we have $\left. D_A \right| = \left. \nabla_A \right|$ and
therefore, to order $\theta=0$, we have
\begin{equation}
\left. D_{\dot A} \left[\frac{1}{\overline R}\left(
\overline{\nabla}^2 +\frac{1}{3} {\overline R} \right) \right]
\right|= \left. \nabla_{\dot A} \left[\frac{1}{\overline R}\left(
\overline{\nabla}^2 +\frac{1}{3} {\overline R} \right) \right]
\right|=0.
\end{equation}
A "rigid" or "curved" superfield whose $\theta=0$ component
vanishes in any frame is identically zero (for a proof see
\cite{Van Nieuwenhuizen:1981ae}). Therefore, we conclude that we
have $D_{\dot A} \left[\frac{1}{\overline R}\left(
\overline{\nabla}^2 +\frac{1}{3} {\overline R} \right) \right]
=0,$ and we may write (\ref{dnabla}).

In the particular gauge (\ref{vielbeinx}), we can write the chiral
density (\ref{phi3}) as
\begin{equation}
\epsilon = \frac{1}{4} \frac{\overline{D}^2 E}{\overline{R}}.
\label{d2e}
\end{equation}
The proof of this fact requires the knowledge of the solution of
the supergravity constraints in terms of unconstrained
superpotentials \cite{gs80}. Indeed, one of these prepotentials is
identical to the chiral compensator. (\ref{d2e}) is obtained from
expressing the supertorsions in terms of the prepotentials
\cite{bk,Gates:1983nr}.

The expansion in components of the chiral density is derived, in the
same gauge, by requiring that $\left. 2\epsilon \right|=e$ and using
its transformation law (\ref{deltaepsilon}) \cite{lr79}. In its
expression, the $\theta$-variables carry Lorentz indices. In these
new $\theta$-variables, the coefficients of the $\theta$-expansion
of chiral superfields are precisely their covariant derivatives
\cite{Muller:1988ux, wessbagger}. A chiral superfield has no
$\overline{\theta}$'s in its expansion. This makes superspace
integration much easier. For ${\cal N}=1, 2$, when we write full
superspace integrals the $\theta$-variables carry Einstein indices,
but when the integrals are in half superspace ($d^2\theta$ in ${\cal
N}=1$, $d^4\theta$ in ${\cal N}=2$), they carry Lorentz indices.
Therefore, one finally has for (\ref{dnabla})
\begin{equation}
{\cal L}= -\frac{3}{4 \kappa^2} \int \epsilon \left[\left(
\overline{\nabla}^2 + \frac{1}{3} \overline{R} \right) \Phi \right]
d^2\theta + \mathrm{h.c.}.
\end{equation}

By writing (\ref{actiong}) on this form, one can identify the
lagrangian of supergravity minimally coupled to a chiral field
\cite{wessbagger, Cremmer:1978hn}. The lagrangian of pure
supergravity is simply obtained by taking $\Phi=1.$

%%%%%%%%%%%%%%%%%%%%%%%%%%%%%%%%%%%%%%%%%%%%%%%%%%%%%%%%%%%%%%%%%%%%%%
%%%%%%%%%%%%%%%%%%%%%%%%%%%%%%%%%%%%%%%%%%%%%%%%%%%%%%%%%%%%%%%%%%%%%%

\section{${\cal N}=2$ supergravity in superspace}

\subsection{${\cal N}=2$ conformal supergravity}

The ${\cal N}=2$ Weyl multiplet has 24+24 degrees of freedom. Its
field content is given by the graviton $e_{\mu}^m$, the gravitinos
$\psi_\mu^{A a}$, the U(2) connection $\widetilde{\Phi}_\mu^{ab}$,
an antisymmetric tensor $W_{mn}$ which we decompose as $W_{A \dot
A B \dot B} = 2 \varepsilon_{\dot A \dot B} W_{AB} +2
\varepsilon_{AB} W_{\dot A \dot B}$, a spinor $\Lambda_A^a$ and,
as auxiliary field, a dimension 2 scalar $I$. In superspace, a
gauge choice can be made (in the supercoordinate transformation)
such that the graviton and the gravitinos are related to
$\theta=0$ components of the supervielbein (symbolically $\left.
E_\Pi^{\ N }\right| $):
\begin{equation}
\left. E_\Pi ^{\ \ N}\right| =\left[ \begin{array}{ccc} e_\mu^{\
m} & \frac{1}{2}\psi_\mu^{\ Aa } & \frac{1}{2}\psi_\mu^{\ \dot A
a}\\ 0 & -\delta_B^{\ A} \delta_b^{\ a} & 0 \\ 0 & 0 &
-\delta_{\dot B}^{\ \dot A} \delta_b^{\ a}
\end{array} \right]. \label{vielbeinx2}
\end{equation}
In the same way, we gauge the fermionic part of the Lorentz
superconnection at order $\theta=0$ to zero and we can set its
bosonic part equal to the usual spin connection:
\begin{equation}
\left. \Omega_{\mu m}^{\ \ \ n} \right| = \omega_{\mu m}^{\ \ \ n}
\left(e, \psi^a \right), \left. \Omega_{Aa m}^{\ \ \ \ n}\right|,
\left. \Omega_{\dot Aa m}^{\ \ \ \ n} \right| =0.
\label{connectionx2}
\end{equation}
The U(2) superconnection $\widetilde{\Phi}_\Pi^{ab}$ is such that
$\left. \widetilde{\Phi}_\mu^{ab}\right|=
\widetilde{\Phi}_\mu^{ab}.$ The other fields are the $\theta=0$
component of some superfield, which we write in the same way.

The chiral superfield $W_{AB}$ is the basic object of ${\cal N}=2$
conformal supergravity, in terms of which its action is written.
Other theories with different ${\cal N}$ have an analogous
superfield (e.g. $W_{ABC}$ in ${\cal N}=1$).

In U(2) ${\cal N}=2$ superspace there is an off-shell solution to
the Bianchi identities. The torsions and curvatures can be
expressed in terms of superfields $W_{AB}$, $Y_{AB}$, $U_{A \dot
A}^{ab}$, $X_{ab}$, their complex conjugates and their covariant
derivatives. Of these four superfields, only $W_{AB}$ transforms
covariantly under super-Weyl transformations. The other three
superfields transform non-covariantly; they describe all the
non-Weyl covariant degrees of freedom in the transformation
parameter $H$, and can be gauged away by a convenient
(Wess-Zumino) gauge choice. Another nice feature of ${\cal N}=2$
superspace is that there exists, analogously to the ${\cal N}=1$
case, a chiral density $\epsilon$ which allows us to write chiral
actions \cite{muller872}.

%%%%%%%%%%%%%%%%%%%%%%%%%%%%%%%%%%%%%%%%%%%%%%%%%%%%%%%%%%%%%%%%%%%%%%
%%%%%%%%%%%%%%%%%%%%%%%%%%%%%%%%%%%%%%%%%%%%%%%%%%%%%%%%%%%%%%%%%%%%%%
\subsection{Degauging U(1)}

The first step for obtaining the Poincar\'e theory is to couple to
the conformal theory an abelian vector multiplet (with central
charge), described by a vector $A_\mu$, a complex scalar, a
Lorentz-scalar SU(2) triplet and a spinorial SU(2) dublet. The
vector $A_\mu$ is the gauge field of central charge transformations;
it corresponds, in superspace, to a 1-form $A_\Pi$ with a U(1) gauge
invariance (the central charge transformation). This 1-form does not
belong to the superspace geometry. Using the U(1) gauge invariance
we can set the gauge $\left. A_\Pi \right| = \left(A_\mu, 0
\right).$ The field strength $F_{\Pi \Sigma}$ is a two-form defined
as $F_{\Pi \Sigma}=2 D_{\left[\Pi \right.} A_{\left. \Sigma \right
\}}$ or, with flat indices, $F_{M N}=2 \nabla_{\left[M \right.}
A_{\left. N \right \}} +T_{MN}^{\ \ \ \ P} A_P.$ It satisfies its
own Bianchi identities $D_{\left[\Gamma \right.} F_{\left.\Pi \Sigma
\right \}}=0$ or, with flat indices,
\begin{equation}
\nabla_{\left[M \right.} F_{\left.NP \right \}} +T_{\left.MN
\right |}^{\ \ \ \ Q} F_{Q\left| P \right \}}=0.
\end{equation}
Here we split the U(2) superconnection $\widetilde{\Phi}_\Pi^{ab}$
into a SU(2) superconnection $\Phi_\Pi^{ab}$ and a U(1)
superconnection $\varphi_\Pi$; only the later acts on $A_\Pi$:
$\widetilde{\Phi}_\Pi^{ab}=\Phi_\Pi^{ab} -\frac{1}{2}
\varepsilon^{ab} \varphi_\Pi.$ One has to impose covariant
constraints on its components (like in the torsions), in order to
construct invariant actions:
\begin{equation}
F_{AB}^{ab}= 2 \sqrt{2} \varepsilon_{AB} \varepsilon^{ab} F, \,\,
F_{A \dot B}^{ab} = 0.
\end{equation}
By solving the $F_{MN}$ Bianchi identities with these constraints,
we conclude that they define an off-shell ${\cal N}=2$ vector
multiplet, given by the $\theta=0$ components of the superfields
$A_\mu, F, F^a_A=\frac{i}{2} F^{\dot A a}_{\ \ \ A \dot A},
F^a_b=\frac{1}{2} \left(-\nabla^B_b F^a_B +F \overline{X}^a_{\ b}+
\overline{F} X^a_{\ b} \right).$ $\left. F^a_b \right|$ is an
auxiliary field; $F^a_a=0$ if the multiplet is abelian (as it has to
be in this context). $\overline{F}$ is a Weyl covariant chiral
superfield, with nonzero U(1) and Weyl weigths. A superconformal
chiral lagrangian for the vector multiplet is
\begin{equation}
{\cal L}= \int \overline{\epsilon} F^2 d^4 \overline{\theta} +
\mathrm{h.c.}.
\end{equation}
In order to get a Poincar\'e theory, we must break the
superconformal and local abelian (from the U(1) subgroup of U(2) -
not the gauge invariance of $A_\mu$) invariances. For that, we set
the Poincar\'e gauge $F=\overline{F}=1.$ As a consequence, from
the Bianchi and Ricci identities we get
\begin{equation}
\varphi_A^a=0, \,\, F_a^A = 0.
\end{equation}
Furthermore, $U_{A \dot A}^{ab}$ is an SU(2) singlet, to be
identified with the bosonic U(1) connection (now an auxiliary
field):
\begin{equation}
U_{A \dot A}^{ab} = \varepsilon^{ab} U_{A \dot A} =
\varepsilon^{ab} \varphi_{A \dot A}.
\end{equation}\
Other consequences are
\begin{eqnarray}
F_{A \dot A B \dot B} &=& \sqrt{2} i \left[ \varepsilon_{AB}
\left(W_{\dot A \dot B} + Y_{\dot A \dot B} \right) +
\varepsilon_{\dot A \dot B} \left(W_{AB} + Y_{AB} \right) \right],
\label{fyw} \\ F^a_b &=& X^a_b, \label{fx}\\ \overline{X_{ab}}
&=&X^{ab}.
\end{eqnarray}
(\ref{fyw}) shows that $W_{mn}$ is now related to the vector field
strength $F_{mn}$. $Y_{mn}$ emerges as an auxiliary field, like
$X_{ab}$ (from (\ref{fx})). We have, therefore, the minimal field
representation of ${\cal N}=2$ Poincar\'e supergravity, with a
local SU(2) gauge symmetry and 32+32 off-shell degrees of freedom:
\begin{equation}
e_{\mu}^m, \psi_\mu^{A a}, A_\mu, \Phi_\mu^{ab}, Y_{mn}, U_m,
\Lambda_A^a, X_{ab}, I. \label{32i}
\end{equation}
Although the algebra closes with this multiplet, it does not admit
a consistent lagrangian because of the higher-dimensional scalar
$I$ \cite{bs81}.

%%%%%%%%%%%%%%%%%%%%%%%%%%%%%%%%%%%%%%%%%%%%%%%%%%%%%%%%%%%%%%%%%%%%%%
%%%%%%%%%%%%%%%%%%%%%%%%%%%%%%%%%%%%%%%%%%%%%%%%%%%%%%%%%%%%%%%%%%%%%%
\subsection{Degauging SU(2)}

The second step is to break the remaining local SU(2) invariance.
This symmetry can be partially broken (at most, to local SO(2))
through coupling to a compensating so-called "improved tensor
multiplet" \cite{wpp83, muller871}, or broken completely. We take
the later possibility. There are still two different versions of
off-shell ${\cal N}=2$ supergravity without SO(2) symmetry, each
with different physical degrees of freedom. In both cases we start
by imposing a constraint on the SU(2) parameter $L^{ab}$ which
restricts it to a compensating nonlinear multiplet \cite{whp80} (at
the linearized level, $\nabla_A^{\underline{a}} L^{\underline{b}
\underline{c}}=0$). From the transformation law of the SU(2)
connection $\delta \Phi_M^{ab}=-\nabla_M L^{ab}$ we can get the
required condition for $L^{ab}$ by imposing the following constraint
on the fermionic connection:
\begin{equation}
\Phi_A^{abc}=2 \varepsilon^{a \underline{b}}
\rho_A^{\underline{c}}. \label{defrho}
\end{equation}
This constraint requires introducing a new fermionic superfield
$\rho_A^a$. We also introduce its fermionic derivatives $P$ and
$H_m$. The previous SU(2) connection $\Phi_\mu^{ab}$ is now an
unconstrained auxiliary field. The divergence of $H_m$ is
constrained, though, at the linearized level by the condition
$\nabla^m H_m=\frac{1}{3} R-\frac{1}{12} I,$ which is equivalent to
saying that $I$ is no longer an independent field. This constraint
implies that only the transverse part of $H_m$ belongs to the
nonlinear multiplet; its divergence lies in the original Weyl
multiplet. From the structure equation (\ref{cur}) and the
definition (\ref{defrho}), we can derive off-shell relations for the
(still SU(2) covariant) derivatives of $\rho^a_A$. Altogether, these
component fields form then the "old minimal" ${\cal N}=2$ 40+40
multiplet \cite{fv79}: $e_{\mu}^m, \psi_\mu^{A a}, A_\mu,
\Phi_\mu^{ab}, Y_{mn}, U_m, \Lambda_A^a, X_{ab}, H_m, P, \rho_A^a.$
This is "old minimal" ${\cal N}=2$ supergravity, the formulation we
are working with. The final lagrangian can be found in
\cite{whp80,muller84}. The other possibility (also with SU(2)
completely broken) is to further restrict the compensating
non-linear multiplet to an on-shell scalar multiplet
\cite{muller86}. This reduction generates a minimal 32+32 multiplet
(not to be confused with (\ref{32i})) with new physical degrees of
freedom. We will not further pursue this version of ${\cal N}=2$
supergravity.
%%%%%%%%%%%%%%%%%%%%%%%%%%%%%%%%%%%%%%%%%%%%%%%%%%%%%%%%%%%%%%%%%%%%%%
%%%%%%%%%%%%%%%%%%%%%%%%%%%%%%%%%%%%%%%%%%%%%%%%%%%%%%%%%%%%%%%%%%%%%%

\subsection{From ${\cal N}=2$ SU(2) superspace to $x-$space} \label{appendix2}

Our choices for torsion constraints in ${\cal N}=2$ are very similar
to the ones for generic ${\cal N}$ presented in section \ref{cc},
the only difference being that, like in ${\cal N}=1$, we have the
representation-preserving constraints $T_{A B}^{a b m}, T_{A a B b
\dot C c}= 0$. In conformal supergravity, all torsions and
curvatures can be expressed in terms of the basic superfields
$W_{AB}$, $Y_{AB}$, $U_{A \dot A}$, $X_{ab}$. After breaking of
superconformal invariance and local U(2), the basic superfields in
the Poincar\'e theory become the physical field $W_{AB}$ and the
auxiliary field $\rho_A^a$ \cite{gates80}. All torsions and
curvatures can be expressed off-shell in terms of these superfields,
their complex conjugates and derivatives \cite{muller84}. $\left.
W_{AB} \right|$, at the linearized level, is related to the field
strength of the physical vector field $A_\mu$ (the graviphoton): from (\ref{fyw}),
\begin{equation}
\left. W_{AB} \right|= -\frac{i}{2 \sqrt{2}} \sigma^{mn}_{AB}
F_{mn} -Y_{AB} -\frac{i}{4} \sigma^{mn}_{AB} \left(\psi_m^{Cc}
\psi_{nCc} +\psi_m^{\dot Cc} \psi_{n \dot Cc} \right).
\label{fyw2}
\end{equation}
$X^{ab}=\frac{1}{2} \left( \nabla^{\dot A \underline{a}}
-2\rho^{\dot A \underline{a}} \right) \rho_{\dot
A}^{\underline{b}},$ $Y_{AB}= -\frac{i}{2} \left(
\nabla^a_{\underline{A}}+2 \rho^a_{\underline{A}} \right)
\rho_{\underline{B} a},$ $H_{A \dot A}= -i \nabla_A^a \rho_{\dot
A a} +i \nabla_{\dot A}^a \rho_{A a},$ $P=i \nabla^{\dot A a} \rho_{\dot A a},$
$\Phi_{A \dot A}^{ab}= \frac{i}{2} \left( \nabla_A^{\underline{a}}
\rho_{\dot A}^{\underline{b}} - \nabla_{\dot A}^{\underline{a}}
\rho_A^{\underline{b}} -4 \rho_A^{\underline{a}} \rho_{\dot
A}^{\underline{b}} \right),$ $U_{A \dot A}= \frac{1}{4} \left( \nabla_A^a \rho_{\dot A a}
+\nabla_{\dot A}^a \rho_{A a} +4 \rho_A^a \rho_{\dot A a} \right),$ $\Lambda^{A a}= -i \nabla^A_b
X^{ab}$ are auxiliary fields at $\theta=0$; $I=i \nabla^{\dot A a}
\Lambda_{\dot A a} - i \nabla^{A a} \Lambda_{A a}$ is a dependent
field. In the linearized approximation,
\begin{align}
\left. W_{B C A a}\right|&= \frac{i}{2}\nabla_{B a} W_{CA}\left.|
-\frac{i}{6} \left(\varepsilon_{BC} \Lambda_{Aa} +
\varepsilon_{BA} \Lambda_{Ca} \right)\right| =-\frac{1}{4}
\psi_{ABCc} +\ldots, \nonumber \\ \left. Y_{B C \dot A a}
\right|&=- \left.\frac{i}{2} \nabla_{\dot A a}Y_{B C} \right|=
-\frac{1}{8}\psi_{BC \dot Aa} +\ldots, \nonumber \\ \left.
W_{ABCD} \right| &= \left. \left(\frac{i}{4}
\nabla_{\underline{A}}^b \nabla_{\underline{B} b} -2
Y_{\underline{AB}} \right) W_{\underline{CD}} \right|
=-\frac{1}{8} {\cal W}^+_{\mu \nu \rho \sigma} \sigma^{\mu
\nu}_{\underline{AB}} \sigma^{\rho \sigma}_{\underline{CD}}
+\ldots, \label{w41} \\ \left. P_{AB \dot A \dot B} \right| &=
\left(\left. \frac{i}{8} \nabla_{\underline{A}}^b
\nabla_{\underline{B} b} Y_{\dot A \dot B} +\mathrm{h.c.} \right)
\right| \ldots = \frac{1}{2} \sigma^\mu_{\underline{A}
\underline{\dot C}} \sigma^\nu_{\underline{B} \underline{\dot D}}
\left({\cal R}_{\mu \nu} -\frac{1}{4} g_{\mu \nu} {\cal R}
\right)\ldots, \nonumber \\ \left. R
\right|&=\left(\frac{i}{4}\left.\nabla^{\dot A a} \nabla^{\dot
B}_a W_{\dot A \dot B}-\frac{1}{4} \nabla^{A a} \nabla_A^b X_{ab}
+\mathrm{h.c.} \right)\right| +\ldots= -{\cal R} +\ldots \nonumber
\end{align}

%%%%%%%%%%%%%%%%%%%%%%%%%%%%%%%%%%%%%%%%%%%%%%%%%%%%%%%%%%%%%%%%%%%%%%
\subsection{The chiral density and the chiral projector}

The action of ${\cal N}=2, d=4$ Poincar\'e supergravity is written
in superspace as
\begin{equation}
{\cal L}_{SG} = -\frac{3}{4 \kappa^2} \int \overline{\epsilon} d^4
\overline{\theta} + \mathrm{h.c.}.
\end{equation}
The expansion of the chiral density $\epsilon$ in components, which
allows us to write chiral actions, can be seen in \cite{muller84}.
From the solution to the Bianchi identities one can check that the
following object is an antichiral projector \cite{Muller:1988ux}:
\begin{equation}
\nabla^{Aa} \nabla_A^b \left(\nabla_a^B \nabla_{Bb} +16 X_{ab}
\right) - \nabla^{Aa} \nabla_a^B \left(\nabla_A^b \nabla_{Bb} -16i
Y_{AB} \right).
\end{equation}
When one acts with this projector on any scalar superfield, one
gets an antichiral superfield (with the exception of $W_{AB}$,
only scalar chiral superfields exist in curved ${\cal N}=2$
superspace; other types of chiral superfields are incompatible
with the solution to the Bianchi identities). Together with
$\epsilon$, this projector allows us to write more general actions
in superspace.

%%%%%%%%%%%%%%%%%%%%%%%%%%%%%%%%%%%%%%%%%%%%%%%%%%%%%%%%%%%%%%%%%%%%%%
%%%%%%%%%%%%%%%%%%%%%%%%%%%%%%%%%%%%%%%%%%%%%%%%%%%%%%%%%%%%%%%%%%%%%%

\section{Superstring $\a^3$ effective actions and ${\cal R}^4$ terms in $d=4$}

In $d=4$, there are only two independent real scalar polynomials
made from four powers of the Weyl tensor \cite{Fulling:1992vm},
given by
\begin{eqnarray}
{\cal W}_+^2 {\cal W}_-^2 &=& {\cal W}^{ABCD} {\cal W}_{ABCD}
{\cal W}^{\dot A \dot B \dot C \dot D} {\cal W}_{\dot A \dot B
\dot C \dot D}, \label{r441}\\ {\cal W}_+^4+{\cal W}_-^4 &=&
\left({\cal W}^{ABCD} {\cal W}_{ABCD}\right)^2 + \left({\cal
W}^{\dot A \dot B \dot C \dot D} {\cal W}_{\dot A \dot B \dot C
\dot D}\right)^2. \label{r442}
\end{eqnarray}

We now write the effective actions for type IIB, type IIA and
heterotic superstrings in $d=4$, after compactification from
$d=10$ in an arbitrary manifold, in the Einstein frame
(considering only terms which are simply powers of the Weyl
tensor, without any other fields except their couplings to the
dilaton, and introducing the $d=4$ gravitational coupling constant
$\kappa$):
\begin{align}
\left. \frac{\kappa^2}{e} {\mathcal L}_{\mathrm{IIB}}
\right|_{{\cal R}^4} &= - \frac{\zeta(3)}{32} e^{-6 \phi} \a^3
{\cal W}_+^2 {\cal W}_-^2 - \frac{1}{2^{11} \pi^5} e^{-4 \phi}\a^3
{\cal W}_+^2 {\cal W}_-^2, \label{2bea4} \\ \left.
\frac{\kappa^2}{e} {\mathcal L}_{\mathrm{IIA}} \right|_{{\cal
R}^4} &= - \frac{\zeta(3)}{32} e^{-6 \phi} \a^3 {\cal W}_+^2
{\cal W}_-^2 \nonumber \\ &- \frac{1}{2^{12} \pi^5} e^{-4
\phi}\a^3 \left[\left({\cal W}_+^4 + {\cal W}_-^4 \right) +224
{\cal W}_+^2 {\cal W}_-^2 \right], \label{2aea4}
\\ \left. \frac{\kappa^2}{e} {\mathcal L}_{\mathrm{het}}
\right|_{{\cal R}^2 + {\cal R}^4} &= -\frac{1}{16} e^{-2 \phi} \a
\left({\cal W}_+^2 + {\cal W}_-^2 \right) +\frac{1}{64} \left(1-2
\zeta(3) \right) e^{-6 \phi} \a^3 {\cal W}_+^2 {\cal W}_-^2
\nonumber \\ &- \frac{1}{3\times2^{12} \pi^5} e^{-4 \phi}\a^3
\left[\left({\cal W}_+^4 + {\cal W}_-^4 \right) +20 {\cal W}_+^2
{\cal W}_-^2 \right]. \label{hea4}
\end{align}
These are only the moduli-independent ${\cal R}^4$ terms from
these actions. Strictly speaking not even these terms are
moduli-independent, since they are all multiplied by the volume of
the compactification manifold, a factor we omitted for simplicity.
But they are always present, no matter which compactification is
taken. The complete action, for every different manifold, includes
many other moduli-dependent terms which we do not consider here:
we are mostly interested in a ${\mathbb T}^6$ compactification.

At string tree level, for all these theories in $d=4$ only ${\cal
W}_+^2 {\cal W}_-^2$ shows up. Because of its well known $d=10$
SL$(2,{\mathbb Z})$ invariance, in type IIB theory only the
combination ${\cal W}_+^2 {\cal W}_-^2$ is present in the $d=4$
effective action (\ref{2bea4}). In the other theories, ${\cal
W}_+^4 + {\cal W}_-^4$ shows up at string one loop level. For type
IIA, the reason is the difference between the left and right
movers in the relative GSO projection at one string loop, because
of this theory being nonchiral. Heterotic string theories have
${\mathcal N}=1$ supersymmetry in ten dimensions, which allows
corrections to the sigma model already at order $\a$, including
${\mathcal R}^2$ corrections (forbidden in type II theories in
$d=10$). Because of cancellation of gravitational anomalies,
another ${\cal R}^4$ contribution is needed in heterotic theories,
which when reduced to $d=4$ gives rise to (\ref{r441}) and
(\ref{r442}).

Next we consider the supersymmetrization of these ${\cal R}^4$
terms in $d=4.$
%%%%%%%%%%%%%%%%%%%%%%%%%%%%%%%%%%%%%%%%%%%%%%%%%%%%%%%%%%%%%%%%%%%%%%
%%%%%%%%%%%%%%%%%%%%%%%%%%%%%%%%%%%%%%%%%%%%%%%%%%%%%%%%%%%%%%%%%%%%%%

\subsection{${\mathcal N}=1, 2$ supersymmetrization of ${\cal W}_+^2 {\cal
W}_-^2$}

The supersymmetrization of the square of the Bel-Robinson tensor
${\cal W}_+^2 {\cal W}_-^2$ has been known for a long time, in
simple \cite{Moura:2001xx, Deser:1977nt} and extended
\cite{Deser:1978br,Moura:2002ip} four dimensional supergravity.

%%%%%%%%%%%%%%%%%%%%%%%%%%%%%%%%%%%%%%%%%%%%%%%%%%%%%%%%%%%%%%%%%%%%%%

\subsubsection{${\mathcal N}=1$}

In ${\cal N}=1,$ the lagrangian to be considered is ($\alpha$ is a
numerical constant)
\begin{equation}
{\cal L}_{SG}+{\cal L}_{{\cal R}^4}=\frac{1}{2 \kappa^2} \int
E\left( 1+\alpha \kappa^6 W^2\overline{W}^2\right) d^4\theta.
\label{action}
\end{equation}
From (\ref{dw0}) and (\ref{dw0dot}), the $\alpha$ term represents
the supersymmetrization of ${\cal W}_+^2 {\cal W}_-^2.$ To compute
the variation of this action, we obviously need the constrained
variation of $W_{ABC}$. The details of this calculation are
presented in \cite{Moura:2001xx}, and so is the final result for
$\int \delta \left[ E\left( 1+\alpha \kappa^6
W^2\overline{W}^2\right) \right] d^4\theta$, which we do not
reproduce here again. From this result, the $R, \overline{R}$
field equations are given by
\begin{equation}
R=6\alpha \kappa^6 \frac{\overline{W}^2 \nabla^2 W^2}{1-2\alpha
\kappa^6 W^2 \overline{W}^2}=6\alpha \kappa^6 \overline{W}^2
\nabla^2 W^2+12\alpha^2 \kappa^{12} \overline{W}^4 W^2 \nabla^2 W^2.
\label{r}
\end{equation}
From (\ref{wb203}), we can easily determine $\nabla^n G_n$. This
way, auxiliary fields belonging to the compensating chiral
multiplet can be eliminated on-shell. This is not the case for the
auxiliary fields which come from the Weyl multiplet ($A_m$), as we
obtained, also in \cite{Moura:2001xx}, a complicated differential
field equation for $G_m$.

%%%%%%%%%%%%%%%%%%%%%%%%%%%%%%%%%%%%%%%%%%%%%%%%%%%%%%%%%%%%%%%%%%%%%%

\subsubsection{${\mathcal N}=2$}

Analogously to ${\mathcal N}=1$, we write the ${\mathcal N}=2$
supersymmetric ${\cal R}^4$ lagrangian in superspace, using the
chiral projector and the chiral density, as a correction to the pure
supergravity lagrangian \cite{Moura:2002ip} ($\alpha$ is again a
numerical constant):
\begin{eqnarray}
{\cal L}_{SG}+{\cal L}_{{\cal R}^4}&=& \int \overline{\epsilon}
\left[-\frac{3}{4 \kappa^2} +\alpha \kappa^4 \left(\nabla^{Aa}
\nabla_A^b \left(\nabla_a^B \nabla_{Bb} +16 X_{ab} \right) \right.
\right. \nonumber \\ &-& \left. \left. \nabla^{Aa} \nabla_a^B
\left(\nabla_A^b \nabla_{Bb} -16i Y_{AB} \right) \right)
W^2\overline{W}^2 \right] d^4 \overline{\theta} + \mathrm{h.c.}.
\label{action2chiral}
\end{eqnarray}
From the component expansion (\ref{w41}), the $\alpha$ term
clearly contains $e {\cal W}_+^2 {\cal W}_-^2$.

At this point we proceed with the calculation of the components of
(\ref{action2chiral}) and analysis of its field content. For that,
we use the differential constraints from the solution to the
Bianchi identities and the commutation relations. The process is
straightforward but lengthy \cite{Moura:2002ip}. The results can
be summarized as follows: with the correction
(\ref{action2chiral}), auxiliary fields $X_{ab}$, $\Lambda_{\dot C
c}$, $Y_{\dot A \dot B}$, $U_m$ and $\Phi_m^{ab}$ get derivatives,
and the same should be true for their field equations; therefore,
these superfields cannot be eliminated on-shell. We also fully
checked that superfields $P$ and $H_m$ do not get derivatives
(with the important exception of $\nabla^m H_m$) and, therefore,
have algebraic field equations which should allow for their
elimination on shell. The only auxiliary field remaining is
$\rho_A^a$. We did not analyze its derivatives because that would
require computing a big number of terms and, for each term, a huge
number of different contributions. Its derivatives should cancel,
though: otherwise, we would have a field ($\rho_A^a$) with a
dynamical field equation while having two fields obtained from its
spinorial derivatives ($P$ and the transverse part of $H_m$)
without such an equation. $\rho_A^a$, like $P$ and transverse
$H_m$, are intrinsic to the "old minimal" version of ${\cal N}=2$
supergravity; they all belong to the same nonlinear multiplet. The
physical theory does not depend on these auxiliary fields and,
therefore, it seems natural that they can be eliminated from the
classical theory and its higher-derivative corrections.

%%%%%%%%%%%%%%%%%%%%%%%%%%%%%%%%%%%%%%%%%%%%%%%%%%%%%%%%%%%%%%%%%%%%%%
%%%%%%%%%%%%%%%%%%%%%%%%%%%%%%%%%%%%%%%%%%%%%%%%%%%%%%%%%%%%%%%%%%%%%%

\subsection{${\mathcal N}=1$ supersymmetrization of ${\cal W}_+^4 + {\cal
W}_-^4$}

For the term ${\cal W}_+^4 + {\cal W}_-^4$ there is a "no-go
theorem", which goes as follows \cite{Christensen:1979qj}: for a
polynomial $I({\cal W})$ of the Weyl tensor to be
supersymmetrizable, each one of its terms must contain equal
powers of ${\cal W}^+_{\mu \nu \rho \sigma}$ and ${\cal W}^-_{\mu
\nu \rho \sigma}$. The whole polynomial must then vanish when
either ${\cal W}^+_{\mu \nu \rho \sigma}$ or ${\cal W}^-_{\mu \nu
\rho \sigma}$ do.

The derivation of this result is based on ${\mathcal N}=1$
chirality arguments, which require equal powers of the different
chiralities of the gravitino in each term of a superinvariant. The
rest follows from the supersymmetric completion. That is why the
only exception to this result is ${\cal W}^2 = {\cal W}_+^2 +
{\cal W}_-^2$: in $d=4$ this term is part of the Gauss-Bonnet
topological invariant (it can be made equal to it with suitable
field redefinitions). This term plays no role in the dynamics and
it is automatically supersymmetric; its supersymmetric completion
is 0 and therefore does not involve the gravitino.

The derivation of \cite{Christensen:1979qj} has been obtained using
${\mathcal N}=1$ supergravity, whose supersymmetry algebra is a
subalgebra of ${\mathcal N}>1$. Therefore, it should remain valid
for extended supergravity too. But one must keep in mind the
assumptions which were made, namely the preservation by the
supersymmetry transformations of $R$-symmetry which, for ${\mathcal
N}=1$, corresponds to U(1) and is equivalent to chirality. In
extended supergravity theories $R-$symmetry is a global internal
$\mbox{U}\left({\mathcal N}\right)$ symmetry, which generalizes (and
contains) U(1) from ${\mathcal N}=1$.

Preservation of chirality is true for pure ${\mathcal N}=1$
supergravity, but to this theory and to most of the extended
supergravity theories one may add matter couplings and extra terms
which violate U(1) $R$-symmetry and yet can be made
supersymmetric, inducing corrections to the supersymmetry
transformation laws which do not preserve U(1) $R$-symmetry.

Having this in mind \cite{Moura:2007ks}, we consider a chiral
multiplet, represented by a chiral superfield $\mathbf{\Phi}$ (we
could take several chiral multiplets $\Phi_i$, which show up after
$d=4$ compactifications of superstring and heterotic theories and
truncation to ${\mathcal N}=1$ supergravity, but we restrict
ourselves to one for simplicity), and containing a scalar field
$\Phi = \left. \mathbf{\Phi} \right|$, a spin$-\frac{1}{2}$ field
$\left. \nabla_A \mathbf{\Phi} \right|$, and an auxiliary field
$F=-\frac{1}{2} \left. \nabla^2 \mathbf{\Phi} \right|$. This
superfield and its hermitian conjugate couple to ${\mathcal N}=1$
supergravity in its simplest version through a superpotential
\begin{equation}
P\left(\mathbf{\Phi}\right)=d + a \mathbf{\Phi} + \frac{1}{2} m
\mathbf{\Phi}^2 + \frac{1}{3} g \mathbf{\Phi}^3 \label{p}
\end{equation}
and a K\"ahler potential $K\left(\mathbf{\Phi},
\overline{\mathbf{\Phi}} \right)=-\frac{3}{\kappa^2} \ln
\left(-\frac{\Omega\left(\mathbf{\Phi}, \overline{\mathbf{\Phi}}
\right)}{3} \right),$ with
\begin{equation}
\Omega\left(\mathbf{\Phi}, \overline{\mathbf{\Phi}} \right)=-3+
\mathbf{\Phi} \overline{\mathbf{\Phi}} + c \mathbf{\Phi} +
\overline{c} \overline{\mathbf{\Phi}}. \label{o}
\end{equation}

In order to include the term (\ref{r442}), we take the following
effective action:
\begin{eqnarray}
{\cal L}&=&-\frac{1}{6 \kappa^2} \int E
\left[\Omega\left(\mathbf{\Phi}, \overline{\mathbf{\Phi}} \right)
+ \a^3 \left(b \mathbf{\Phi} \left(\nabla^2 W^2\right)^2 +
\overline{b} \overline{\mathbf{\Phi}} \left(\overline{\nabla}^2
\overline{W}^2\right)^2 \right) \right] d^4\theta \nonumber
\\ &-&\frac{2}{\kappa^2} \left(\int \epsilon
P\left(\mathbf{\Phi}\right) d^2\theta + \mathrm{h.c.} \right).
\label{r421}
\end{eqnarray}

If one expands (\ref{r421}) in components, one does not directly
get (\ref{r442}), but one should look at the auxiliary field
sector. Because of the presence of the higher-derivative terms,
the auxiliary field from the original conformal supermultiplet
$A_m$ also gets higher derivatives in its equation of motion, and
therefore it cannot be simply eliminated
\cite{Moura:2001xx,Moura:2002ip}. Because the auxiliary fields $M,
N$ belong to the chiral compensating multiplet, their field
equation should be algebraic, despite the higher derivative
corrections \cite{Moura:2001xx,Moura:2002ip}. That calculation
should still require some effort; plus, those $M, N$ auxiliary
fields should not generate by themselves terms which violate U(1)
$R$-symmetry: these terms should only occur through the
elimination of the chiral multiplet auxiliary fields $F, \bar{F}.$
This is why we will only be concerned with these auxiliary fields,
which therefore can be easily eliminated through their field
equations \cite{Cremmer:1978hn}. The final result, taking into
account only terms up to order $\a^3$, is
\begin{eqnarray}
\kappa^2 {\cal L}_{F, \overline{F}}&=& -15 e \frac{ \left(3 + c
\overline{c}\right)}{\left(3 + 4 c \overline{c}\right)^2} \left(m
\overline{a} \Phi + \overline{m} a \overline{\Phi} \right) \left(c
\Phi + \overline{c} \overline{\Phi} \right) \nonumber \\ &+& e
\frac{2 c^3 \overline{c}^3 + 60 c^2 \overline{c}^2 + 117 c
\overline{c}-135}{\left(3 + 4 c \overline{c}\right)^3} a
\overline{a} \Phi \overline{\Phi} - 36 \a^3 e \left( b
\overline{c} \left(\nabla^2 W^2\right)^2\right| \nonumber \\ &+&
\overline{b} c \left. \left. \left(\overline{\nabla}^2
\overline{W}^2\right)^2 \right| \right) \frac{a \overline{a} + m
\overline{a} \Phi + \overline{m} a \overline{\Phi} + g
\overline{a} \Phi^2 + \overline{g} a \overline{\Phi}^2 + m
\overline{m} \Phi \overline{\Phi} }{\left(3 + 4 c
\overline{c}\right)^2} \nonumber
\\ &-& 3 \a^3 a\overline{a} \frac{74 c^2 \overline{c}^2 + 192 c
\overline{c}-657}{\left(3 + 4 c \overline{c}\right)^4} \Phi
\overline{\Phi} \left( b \overline{c} \left(\nabla^2
W^2\right)^2\right| + \overline{b} c \left. \left.
\left(\overline{\nabla}^2 \overline{W}^2\right)^2 \right| \right)
\nonumber \\ &+& 15 \a^3 e \frac{a \overline{a} + m \overline{a}
\Phi + \overline{m} a \overline{\Phi}}{\left(3 + 4 c
\overline{c}\right)^3} \left[ \right. \nonumber \\ && \left.
\left( c^2 \left(21 + 4 c \overline{c}\right) \Phi + \left(-9 + 6
c \overline{c}\right) \overline{\Phi} \right) \overline{b} \left.
\left(\overline{\nabla}^2 \overline{W}\right)^2\right|+
\mbox{h.c.} \right] + \ldots \label{r42s}
\end{eqnarray}
This way we are able to supersymmetrize ${\cal W}_+^4 + {\cal
W}_-^4$, although we had to introduce a coupling to a chiral
multiplet. Since from (\ref{dw0}) and (\ref{dw0dot}) the factor in
front of ${\cal W}_+^4$ (resp. ${\cal W}_-^4$) in (\ref{r42s}) is
given by $\frac{-144 b \overline{c} a \overline{a}}{\left(3 + 4 c
\overline{c}\right)^2}$ (resp. $\frac{-144 \overline{b} c a
\overline{a}}{\left(3 + 4 c \overline{c}\right)^2}$), for this
supersymmetrization to be effective, the factors $a$ from
$P\left(\Phi\right)$ in (\ref{p}) and $c$ from $\Omega \left(\Phi,
\overline{\Phi}\right)$ in (\ref{o}) (and of course $b$ from
(\ref{r421})) must be nonzero.

%%%%%%%%%%%%%%%%%%%%%%%%%%%%%%%%%%%%%%%%%%%%%%%%%%%%%%%%%%%%%%%%%%%%%%
%%%%%%%%%%%%%%%%%%%%%%%%%%%%%%%%%%%%%%%%%%%%%%%%%%%%%%%%%%%%%%%%%%%%%%

\subsubsection{${\cal W}_+^4 + {\cal W}_-^4$ in extended supergravity}

${\cal W}_+^4 + {\cal W}_-^4$ must also arise in extended $d=4$
supergravity theories, for the reasons we saw, but the "no-go"
result of \cite{Christensen:1979qj} should remain valid, since it
was obtained for ${\mathcal N}=1$ supergravity, which can always
be obtained by truncating any extended theory. For extended
supergravities, the chirality argument should be replaced by
preservation by supergravity transformations of U(1), which is a
part of $R$-symmetry.

${\mathcal N}=2$ supersymmetrization of ${\cal W}_+^4 + {\cal
W}_-^4$ should work in a way similar to what we saw for ${\mathcal
N}=1$. ${\mathcal N}=2$ chiral superfields must be Lorentz and
SU(2) scalars but they can have an arbitrary U(1) weight, which
allows supersymmetric U(1) breaking couplings.

A similar result should be more difficult to implement for
${\mathcal N} \geq 3$, because there are no generic chiral
superfields. Still, there are other multiplets than the Weyl, which
one can consider in order to couple to ${\cal W}_+^4 + {\cal W}_-^4$
and allow for its supersymmetrization. The only exception is
${\mathcal N}=8$ supergravity, a much more restrictive theory
because of its higher amount of supersymmetry. In this case one can
only take its unique multiplet, which means there are no extra
matter couplings one can consider. We have
shown that the ${\mathcal N}=8$ supersymmetrization of ${\cal W}_+^4
+ {\cal W}_-^4,$ coupled to scalar fields from the Weyl multiplet,
is not allowed even at the linearized level \cite{Moura:2007ac}. In ${\mathcal N}=8$ superspace one
can only have SU(8) invariant terms, and we argued ${\cal W}_+^4 +
{\cal W}_-^4$ should be only $\mbox{SU}(4) \otimes \mbox{SU}(4)$
invariant. If that is the case, in order to supersymmetrize this
term besides the supergravity multiplet one must introduce
$U-$duality multiplets, with massive string states and
nonperturbative states. The fact that one cannot supersymmetrize in ${\mathcal N}=8$ a
term which string theory requires to be supersymmetric, together
with the fact that one needs to consider nonperturbative states
(from $U-$duality multiplets) in order to understand a
perturbative contribution may be seen as indirect evidence that
${\mathcal N}=8$ supergravity is indeed in the swampland \cite{Green:2007zzb}. We believe that topic deserves
further study.

%%%%%%%%%%%%%%%%%%%%%%%%%%%%%%%%%%%%%%%%%%%%%%%%%%%%%%%%%%%%%%%%%%%%%%
%%%%%%%%%%%%%%%%%%%%%%%%%%%%%%%%%%%%%%%%%%%%%%%%%%%%%%%%%%%%%%%%%%%%%%
\section{Applications to black holes in string theory}
String-corrected black holes have been a very active recent topic
of research, for which one needs to know the string effective
actions to a certain order in $\a.$ Topics which have been studied
include finding $\a-$corrected black hole solutions by themselves,
but also studying their properties like the entropy. One of the biggest
successes of string theory was the calculation of the microscopic entropy
of a class of supersymmetric black holes and the verification that this
result corresponds precisely to the macroscopic result of Bekenstein and
Hawking. Clearly it is very important to find out if and how
this correspondence extends to the full string effective action,
without $\a$ corrections.

Because of different $\a$ corrections each quantity gets, typically the
entropy does not equal one quarter of the horizon area for black holes
with higher derivative terms. In order to compute the entropy for these
black holes, a formula has been developed by Wald \cite{w93}. When this
formula is applied to extremal (not necessarily supersymmetric) black
holes, one arrives at the entropy functional formalism developed by Sen
(for a complete review see \cite{Sen:2007qy}). This formalism can be
summarized as follows: one considers a black hole solution from a
lagrangian $\mathcal{L}$ with gravity plus some gauge fields and massless scalars in
$d$ dimensions. The near horizon limit of such black hole corresponds to
$AdS_2 \times S^{d-2}$ geometry, with two parameters $v_1, v_2.$ Also
close to the horizon, the gauge fields are parameterized by sets of
electric $(e_i)$ and magnetic $(p_a)$ charges, and the scalar fields by
constants $u_s.$ The parameters $\left(\vec{u}, \vec{v}, \vec{e}, \vec{p}\right)$
are up to now arbitrary and, therefore, the solution is off-shell. Next
we define the function (to be evaluated in the near horizon limit)
$$f\left(\vec{u}, \vec{v}, \vec{e}, \vec{p}\right)=\int_{S^{d-2}} \sqrt{-g} \mathcal{L}\,d\Omega_{d-2}.$$
The on-shell values of $\vec{u}, \vec{v}, \vec{e}$ for a given theory
are found through the relations
$$\frac{\partial f}{\partial u_s}=0, \, \frac{\partial f}{\partial v_j}=0, \, \frac{\partial f}{\partial e_i}=q_i,$$
which also reproduce the equations of motion. Then, using Wald's
formulation, Sen derived the black hole entropy, given by
$$S=2 \pi \left( e_i \frac{\partial f}{\partial e_i} - f \right).$$
This process has been verified for extremal (supersymmetric or not) black holes
in generic $d$ dimensions. In particular, it has been tested with off
shell formulations of supergravity \cite{Sahoo:2006rp} (these formulations
are known for ${\mathcal N}=1$ in $d=4$ or ${\mathcal N}=2$ in $d=4,5,6$).
When one considers black holes in these theories, auxiliary fields must also be considered in $f,$ necessarily as
independent fields (since for this functional we take an a-priori off-shell solution).
As we have seen, when considering theories with higher-derivative corrections, some of
these auxiliary fields can still be eliminated, but others become dynamical. Clearly a precise
knowledge of the behavior of the different auxiliary fields, like we have studied, is
essential if one wishes to determine the higher-derivative corrections to black hole
properties such as the entropy.

A particularly well studied case \cite{Lopes Cardoso:1998wt} (which has been reviewed in this volume
\cite{Mohaupt:2008gt}) is that of BPS black holes in
$d=4, {\mathcal N}=2$ supergravity coupled to $n$ vector multiplets,
to which are associated $n$ scalar fields $X^I$ and $n$ vector fields
$A_\mu^I.$ The holomorphic higher-derivative corrections associated to
these black holes are given as higher genus contributions to the
prepotential, in the form of a function
\begin{equation}
F(X^I,\hat{A})=\sum_{g=0}^\infty F^{(g)}(X^I) \hat{A}^g, \label{f}
\end{equation}
$\hat{A}$ being a scalar field which, in our conventions, is given by
$\hat{A}=\left. W^{AB} W_{AB} \right|.$ From (\ref{fyw2}), one sees
that $\hat{A}$ is related to the square of the selfdual part of the
graviphoton field strength $F_{\mu\nu}$, but also to the square of
the auxiliary field $Y_{AB}$ (which, as we saw, may become dynamical
in the presence of higher-derivative terms). From (\ref{w41}), one
immediately sees that a lagrangian containing $F(X^I,\hat{A})$ as an
$F-$term includes ${\cal W}^2$ terms, each multiplied by terms
depending on moduli and on powers of either $F_{\mu\nu}$ or $Y_{mn}.$
These $Y_{mn}$ factors may generate terms with higher powers of the
Weyl tensor ${\cal W}_{\mu\nu\rho\sigma}.$

After some rescaling (in order to have manifest symplectic covariance),
$\hat{A}$ becomes the variable $\Upsilon$, which at the horizon takes
a particular numerical value ($\Upsilon=-64$ in the conventions of
\cite{Mohaupt:2008gt}). This value is universal, independent of the
model taken (i.e. for any function $F(X^I,\hat{A})$ of the form
(\ref{f})), as long as the black hole solution under consideration
is supersymmetric. There may exist other near-horizon configurations
(corresponding to nonsupersymmetric black holes) which extremize the
entropy function but correspond to different attractor equations and
different values for $\Upsilon$. These values are not universal: each
solution has its own (constant) $\Upsilon$.

The generalized prepotential (\ref{f}) does not represent the full set
of higher derivative corrections one must consider in a supersymmetric
theory in $d=4,$ even for a black hole solution. There are also the
nonholomorphic corrections, which are necessary for the entropy to be
invariant under string dualities, as discussed in \cite{Mohaupt:2008gt}.
At the time, the way to incorporate these corrections into the attractor
mechanism is still under study. On general grounds, if $\Upsilon$ is
coupled to the nonholomorphic corrections, then it should in principle
get a different value. This (still unknown) different value for
$\Upsilon$ should also in principle depend on the model which we are taking.
Because of this nonuniversality, we cannot simply take a general expression
for the nonholomorphic corrections: we really need each term,
to the order we are working, in the effective action. For that, in the
cases when auxiliary fields (namely $\Upsilon$) exist and are part of the
higher derivative correction terms (as studied in
\cite{Lopes Cardoso:1999ur}), we must know exactly their behavior in the
presence of such corrections, in the way we presented on the first part of these notes.

%%%%%%%%%%%%%%%%%%%%%%%%%%%%%%%%%%%%%%%%%%%%%%%%%%%%%%%%%%%%%%%%%%%%%%
%%%%%%%%%%%%%%%%%%%%%%%%%%%%%%%%%%%%%%%%%%%%%%%%%%%%%%%%%%%%%%%%%%%%%%
\section{Summary and discussion}

We computed the ${\mathcal R}^4$ terms in the superstring
effective actions in four dimensions. We showed that besides the
usual square of the Bel-Robinson tensor ${\cal W}_+^2 {\cal
W}_-^2,$ the other possible ${\mathcal R}^4$ term in $d=4,$ ${\cal
W}_+^4 + {\cal W}_-^4,$ was also part of two of those actions at
one string loop. We then studied their supersymmetrization.

For ${\cal W}_+^2 {\cal W}_-^2$ we wrote down its
supersymmetrization directly in ${\mathcal N}=1$ and ${\mathcal
N}=2$ superspace, taking advantage of the off-shell formulation of
these theories. The terms we wrote down were off-shell; in both
cases we tried to obtain the on-shell action by eliminating the
auxiliary fields. We noticed that some auxiliary fields could be
eliminated, while others couldn't.

A careful analysis shows that, in both cases we studied, the
auxiliary fields that can be eliminated in the supersymmetrization
of ${\cal W}_+^2 {\cal W}_-^2$ come from multiplets which,
on-shell, have no physical fields; while the auxiliary fields that
get derivatives come from multiplets with physical fields on-shell
(the graviton, the gravitino(s) and, in ${\cal N}=2$, the vector).
Our general conjecture for supergravity theories with higher
derivative terms, which is fully confirmed in the "old minimal"
${\mathcal N}=1, 2$ cases with ${\cal W}_+^2 {\cal W}_-^2$, can
now be stated: the auxiliary fields which come from multiplets
with on-shell physical fields cannot be eliminated, but the ones
that come from compensating multiplets that, on shell, have no
physical fields, can. In order to get more evidence for it, the
analysis we made should also be extended to the other different
versions of these supergravity theories, and with other higher
derivative terms.

We moved on to try to supersymmetrize ${\cal W}_+^4 + {\cal W}_-^4$,
but we faced a previous result stating that supersymmetrization
could not be achieved because in ${\mathcal N}=1$ it would violate
chirality, which is preserved in pure supergravity. The way we found
to circumvent this problem was to couple ${\cal W}_+^4 + {\cal
W}_-^4$ to a chiral multiplet and, after eliminating its auxiliary
fields, obtain that same term on-shell. We worked this out in
${\mathcal N}=1$ supergravity and the same should be possible in
${\mathcal N}=2.$ For ${\mathcal N}=8$ that should not be possible
any longer, because there are no other multiplets we could use to
couple to ${\cal W}_+^4 + {\cal W}_-^4$ that could help us: the Weyl
multiplet is the only one allowed in this theory. This is a sign that
${\mathcal N}=8$ supergravity is indeed in the swampland.

We ended by discussing applications of these results to black holes in
string theory, namely the attractor mechanism and the calculation of
the black hole entropy in the presence of higher derivative terms. We
considered extremal black holes in $d$ dimensions, through Sen's
entropy functional formalism, and in particular BPS black holes in
$d=4, {\mathcal N}=2$ supergravity. In all cases we concluded that,
having those applications in mind, when auxiliary fields exist, one
needs to know exactly their behavior in the presence of such higher
derivative corrections.

%%%%%%%%%%%%%%%%%%%%%%%%%%%%%%%%%%%%%%%%%%%%%%%%%%%%%%%%%%%%%%%%%%%%%%
\paragraph{Acknowledgments}
I thank Ashoke Sen for correspondence and Gabriel Lopes Cardoso for discussions.
I also thank the organizers for the opportunity to participate in an
excellent conference and present this work, which has been partially supported
by Funda\c c\~ao para a Ci\^encia e a Tecnologia through fellowship
BPD/14064/2003 and by FCT and EU FEDER through PTDC via QSec
PTDC/EIA/67661/2006 project.

%\paragraph{} I apologize for the omissions in the list of references, which is necessarily incomplete due to space constraints.
%%%%%%%%%%%%%%%%%%%%%%%%%%%%%%%%%%%%%%%%%%%%%%%%%%%%%%%%%%%%%%%%%%%%%%
%%%%%%%%%%%%%%%%%%%%%%%%%%%%%%%%%%%%%%%%%%%%%%%%%%%%%%%%%%%%%%%%%%%%%%


\begin{thebibliography}{99.}
\bibitem{Van Nieuwenhuizen:1981ae} P.~Van Nieuwenhuizen: Phys.\ Rept. {\bf 68} (1981) 189.
%%CITATION = PRPLC,68,189;%%
\bibitem{dragon79} N. Dragon: Zeit. f\"ur Phys. {\bf C2} (1979), 29.
\bibitem{wz781} J. Wess, B. Zumino: Phys. Lett. {\bf 74B} (1978), 51.
\bibitem{ht78} P. Howe, R.W. Tucker: Phys. Lett. {\bf 80B} (1978),
138.
\bibitem{howe82} P. Howe: Nucl. Phys. {\bf B199} (1982), 309.
\bibitem{Muller:1988ux} M.~M\"uller: \emph{Consistent Classical Supergravity
Theories}, Lect.\ Notes Phys. {\bf 336} (Springer-Verlag, Berlin
Heidelberg New York 1989).
%%CITATION = LNPHA,336,1;%%
\bibitem{muller862} M. M\"uller: Zeit. f\"ur Phys. {\bf C31} (1986), 321.
%%CITATION = ZEPYA,C31,321;%%
\bibitem{gwz79} R. Grimm, J. Wess, B. Zumino: Nucl. Phys.
{\bf B152} (1979), 255.
\bibitem{bk} I.L. Buchbinder, S.M. Kuzenko: \emph{Ideas and Methods of
Supersymmetry and Supergravity} (Institute of Physics, Bristol
1995).
\bibitem{Gates:1983nr} S.~J.~Gates, M.~T.~Grisaru, M.~Ro\v{c}ek,
W.~Siegel: {\it Superspace},\ Front.\ Phys.\ {\bf 58}
(Benjamin-Cummings, Reading 1983) [arXiv:hep-th/0108200].
%%CITATION = FRPHA,58,1;%%
\bibitem{sw78} K. Stelle, P. West: Phys. Lett. {\bf 74B} (1978), 330.
\bibitem{fvn78} S. Ferrara, P. van Nieuwenhuizen: Phys. Lett.
{\bf 74B} (1978), 333.
\bibitem{s79} W. Siegel: Phys. Lett. {\bf 80B} (1979), 224;
P. Breitenlohner: Phys. Lett. {\bf 67B} (1977), 49; idem: Nucl.
Phys. {\bf B124} (1977), 500.
\bibitem{sw81} M.F. Sohnius, P. West: Phys. Lett. {\bf 105B} (1981),
353; idem: {\it Nucl. Phys.} {\bf B198} (1982), 493.
\bibitem{gs80} S. J. Gates, Jr., W. Siegel: Nucl. Phys. {\bf B147}
(1979), 77; idem: {\it Nucl. Phys.} {\bf B163} (1980), 519.
\bibitem{os78} V. Ogievetsky, E. Sokatchev: Phys. Lett {\bf 79B}
(1978), 222.
\bibitem{lr79} M. Ro\v{c}ek, U. Lindstr\"om: Phys. Lett. {\bf 83B}
(1979), 179.
\bibitem{wessbagger} J. Wess, J. Bagger: \emph{Supersymmetry and
Supergravity} (Princeton University Press, Princeton, 1992).
\bibitem{Moura:2001xx} F.~Moura: J.\
High\ Energy\ Phys. {\bf 0109} (2001) 026 [arXiv:hep-th/0106023];
idem: J.\ High\ Energy\ Phys.\ {\bf 0208} (2002) 038
[arXiv:hep-th/0206119].
%%CITATION = HEP-TH 0106023;%%
%%CITATION = HEP-TH 0206119;%%
\bibitem{wz782} J. Wess, B. Zumino: Phys. Lett. {\bf 79B} (1978),
394.
\bibitem{Cremmer:1978hn} E.~Cremmer, B.~Julia et al: Nucl.\ Phys. {\bf B147} (1979)
105.
%%CITATION = NUPHA,B147,105;%%
\bibitem{muller872} M. M\"uller: Nucl. Phys. {\bf B289} (1987), 557.
%%CITATION = NUPHA,B289,557;%%
\bibitem{bs81} P. Breitenlohner, M.F. Sohnius: Nucl. Phys. {\bf B178}
(1981), 151.
\bibitem{wpp83} B. de Wit, R. Philippe, A. van Proeyen: Nucl. Phys. {\bf B219} (1983), 143.
\bibitem{muller871} M. M\"uller: Nucl. Phys. {\bf B282} (1987), 329.
%%CITATION = NUPHA,B282,329;%%
\bibitem{whp80} B. de Wit, J. W. van Holten, A. van Proeyen:
Nucl. Phys. {\bf B167} (1980), 186; idem: Nucl. Phys. {\bf B184}
(1981), 77.
\bibitem{fv79} E.S. Fradkin, M.A. Vasiliev: Lett. Nuovo Cim.
{\bf 25} (1979), 79; idem: Phys. Lett. {\bf 85B} (1979), 47.
\bibitem{muller84} M. M\"uller: Zeit. f\"ur Phys. {\bf C24} (1984), 175.
%%CITATION = ZEPYA,C24,175;%%
\bibitem{muller86} M. M\"uller: Phys. Lett. {\bf 172B} (1986), 353.
%%CITATION = PHLTA,B172,353;%%
\bibitem{gates80} S.J. Gates, Jr.: Nucl. Phys. {\bf B176} (1980), 397; idem: Phys. Lett. {\bf 96B} (1980), 305.
\bibitem{Fulling:1992vm} S.~A.~Fulling, R.~C.~King  et al: Class.\ Quant.\ Grav.\ {\bf 9} (1992) 1151.
%%CITATION = CQGRD,9,1151;%%
\bibitem{Deser:1977nt} S.~Deser, J.~H.~Kay, K.~S.~Stelle:
Phys.\ Rev.\ Lett.\ {\bf 38} (1977) 527.
%%CITATION = PRLTA,38,527;%%
\bibitem{Deser:1978br} S.~Deser and J.~H.~Kay: Phys.\ Lett.\ {\bf B76}
(1978) 400.
%%CITATION = PHLTA,B76,400;%%
\bibitem{Moura:2002ip} F.~Moura: J.\
High\ Energy\ Phys.\ {\bf 0307} (2003) 057 [arXiv:hep-th/0212271].
%%CITATION = HEP-TH 0212271;%%
\bibitem{Christensen:1979qj} S.~Christensen, S.~Deser,
M.~Duff, M.~Grisaru: Phys.\ Lett.\ {\bf B84} (1979) 411.
%%CITATION = PHLTA,B84,411;%%
\bibitem{Moura:2007ks} F.~Moura: J.\ High\
Energy\ Phys.\ {\bf 0706} (2007) 052 [arXiv:hep-th/0703026].
%%CITATION = JHEPA,0706,052;%%
\bibitem{Moura:2007ac}
F.~Moura: Phys.\ Rev.\ {\bf D77} (2008) 125011 [arXiv:0708.3097 [hep-th]].
%%CITATION = PHRVA,D77,125011;%%
\bibitem{Green:2007zzb} M.~B.~Green, H.~Ooguri and J.~H.~Schwarz:
Phys.\ Rev.\ Lett.\ {\bf 99} (2007) 041601 [arXiv:0704.0777 [hep-th]].
%%CITATION = PRLTA,99,041601;%%
\bibitem{w93}
R.~M.~Wald: Phys.\ Rev.\ \textbf{D48} (1993) 3427 [arXiv:gr-qc/9307038].
%%CITATION = GR-QC 9307038;%%
\bibitem{Sen:2007qy} A.~Sen: Gen.\ Rel.\ Grav.\ {\bf 40} (2008) 2249 [arXiv:0708.1270 [hep-th]].
%%CITATION = GRGVA,40,2249;%%
\bibitem{Sahoo:2006rp} B.~Sahoo and A.~Sen: J.\ High\
Energy\ Phys.\ {\bf 0609} (2006) 029 [arXiv:hep-th/0603149].
%%CITATION = JHEPA,0609,029;%%
\bibitem{Lopes Cardoso:1998wt} G.~Lopes Cardoso, B.~de Wit
and T.~Mohaupt: Phys.\ Lett.\ {\bf B451} (1999) 309 [arXiv:hep-th/9812082].
%%CITATION = PHLTA,B451,309;%%
\bibitem{Mohaupt:2008gt} T.~Mohaupt: \emph{From Special Geometry to Black Hole Partition Functions}, this volume, arXiv:0812.4239 [hep-th].
%%CITATION = ARXIV:0812.4239;%%
\bibitem{Lopes Cardoso:1999ur} G.~Lopes Cardoso, B.~de Wit and T.~Mohaupt: Nucl.\ Phys.\ {\bf B567} (2000) 87 [arXiv:hep-th/9906094].
%%CITATION = NUPHA,B567,87;%%
\end{thebibliography}
\end{document}